\documentclass[%
superscriptaddress,
 amsmath,amssymb,
 aps,
prb,
twocolumn,
]{revtex4-1}
\usepackage[italicdiff]{physics}
\usepackage[dvipdfmx]{graphicx}
\usepackage{dcolumn,amsmath,txfonts,here,subfigure}
\usepackage{bm}
\usepackage{color}

\newcommand{\average}[1]{\ensuremath{\langle#1\rangle} }

\begin{document}

\preprint{APS/123-QED}

\title{Universal scrambling in gapless quantum spin chains}

\author{Shunsuke Nakamura}
\affiliation{Department of Applied Physics, The University of Tokyo, \\
	7-3-1 Hongo, Bunkyo-ku, Tokyo 113-8656, Japan
}
\author{Eiki Iyoda}%
\affiliation{Department of Physics, Tokai University, \\
	4-1-1, Kitakaname, Hiratsuka-shi, Kanagawa 259-1292, Japan
}
\author{Tetsuo Deguchi}
\affiliation{Department of Physics, Faculty of Core Research, Ochanomizu University,\\ 2-1-1 Ohtsuka, Bunkyo-ku, Tokyo 112-8610, Japan}
\author{Takahiro Sagawa}
\affiliation{Department of Applied Physics, The University of Tokyo, \\
	7-3-1 Hongo, Bunkyo-ku, Tokyo 113-8656, Japan
}

\date{\today}

\begin{abstract}
Information scrambling, characterized by the out-of-time-ordered correlator (OTOC), has attracted much attention, as it sheds new light on chaotic dynamics in quantum many-body systems. The scale invariance, which appears near the quantum critical region in condensed matter physics, is considered to be important for the fast decay of the OTOC. In this paper, we focus on the one-dimensional spin-$1/2$ XXZ model, which exhibits quantum criticality in a certain parameter region, and investigate the relationship between scrambling and the scale invariance. We quantify scrambling by the averaged OTOC over the Pauli operator basis, which is related to the operator space entanglement entropy (OSEE). Using the infinite time-evolving block decimation (iTEBD) method, we numerically calculate time dependence of the OSEE in the early time region in the thermodynamic limit. We show that the averaged OTOC decays faster in the gapless region than in the gapped region. In the gapless region, the averaged OTOC behaves in the same manner regardless of the anisotropy parameter. This result is consistent with the fact that the low energy excitations of the gapless region belong to the same universality class as the Tomonaga-Luttinger liquid with the central charge $c=1$. Furthermore, we estimate $c$ by fitting the numerical data of the OSEE with an analytical result of the two-dimensional conformal field theory, and confirmed that $c$ is close to unity. Thus, our numerical results suggest that the scale invariance leads to a universal behavior of the OTOC that is independent of the anisotropic parameter, which reflects the universality of the two-dimensional conformal field theory at low temperatures.Although the one-dimensional XXZ model is integrable, our results suggest such a universal behavior of generic non-integrable systems, because the Tomonaga-Luttinger liquid serves as a low-energy effective theory for many non-integrable and integrable systems. On the other hand, the OTOC in our numerical result does not exhibit the exponential decay, as our parameter regime is far from the semiclassical limit.
\end{abstract}

\pacs{Valid PACS appear here}
\maketitle


\section{Introduction}
While classical chaos is characterized by the Lyapunov exponent~\cite{Gallavotti}, the definition of chaos in quantum systems is non-trivial, and many studies have been conducted to understand various aspects of quantum chaos~\cite{Stockmann,Haake}. In addition to the conventional indicators of quantum chaos such as the level statistics~\cite{Stockmann,Haake} and the Loschmidt echo~\cite{Gorin2006}, {\it scrambling} recently attracts attention because it sheds new light on chaotic behaviors in quantum many-body systems~\cite{Hayden2007,Sekino2008}. Scrambling represents the delocalization process of locally encoded quantum information. This is characterized by the out-of-time-ordered correlator (OTOC), which is defined as
\begin{align}
\label{eq:OTOC}
F(t):=\average{\hat{W}(t)\hat{V}(0)\hat{W}(t)\hat{V}(0)}_\beta,
\end{align}
where $\hat{V}$ and $\hat{W}$ are local Hermitian operators, and $\average{\cdots}_\beta$ represents the canonical ensemble average at the inverse temperature $\beta$. We note that the ordering of the times in the OTOC is different from that for the usual correlation functions. As the OTOC is closely related to the square of the commutator $C(t):=-\average{\qty[\hat{W}(t),\hat{V}]^2}$, it quantifies how the operators $W(t)$ and $V(0)$ become non-commutable as the spatial support of $W(t)$ increases in time evolution. In quantum systems, the squared commutator is expected to increase exponentially in the early time regime, if the system scrambles fast enough, which leads to the exponential behavior of the OTOC of the form $F(t)\approx a-b\mathrm{e}^{\lambda_L t}$~\cite{Maldacena2016,Kitaevtalk}. Here, $\lambda_\mathrm{L}$ is referred to as the quantum Lyapunov exponent. In terms of the quantum information theory, the averaged OTOC over the operator basis is equivalent to tripartite mutual information~\cite{Hosur2016} and represents quantum pseudorandomness~\cite{Roberts2017}.

As an important feature of the OTOC, its decay rate has an upper bound that is referred to as the chaotic bound or the Maldacena-Shenker-Stanford (MSS) bound~\cite{Maldacena2016}. This bound is achieved by the Sachdev-Ye-Kitaev (SYK) model~\cite{Kitaevtalk,Sachdev1992,Sachdev2015,Maldacena2016a,Fu2016,Cotler2017b}, which exhibits the partial scale invariance SL(2,$\mathbf{R}$). Moreover, the two-dimensional conformal field theory (2D CFT)~\cite{Francesco} in the large central charge limit also achieves the upper bound~\cite{Roberts2015}.
We emphasize that these two models exhibits the scale invariance.
We note that the role of scale invariance has also been studied with the Tomonaga-Luttinger liquid (TLL)~\cite{Dora2017} and the Bose-Hubbard model~\cite{Shen2017,Bohrdt2017}.

In this paper, we focus on the relationship between the scale invariance and scrambling with a simple quantum spin chain. Specifically, we numerically investigate the one-dimensional quantum spin-$1/2$ anti-ferromagnetic XXZ chain, which exhibits a quantum phase transition between the gapless phase and the gapped phase. The low-energy excitations in the gapless phase is characterized by the TLL and those in the gapped phase are described by the Ising-like excitations~\cite{Giamarchi}. 
For this model, we calculate the averaged OTOC that is equivalent to the operator space entanglement entropy (OSEE)~\cite{Zanardi2001,Prosen2007,Znidaric2008a,Pizorn2009,Xu}.

We show that there is a qualitative difference between the behavior of OTOC in the gapless region and that of gapped region: the decay of OTOC in the gapless region is faster than that of the gapped one. This result supports the expectation that the scale invariance of CFT in a quantum system leads to fast scrambling~\cite{Maldacena2016,Kitaevtalk}. Furthermore, our numerical study shows that the decaying behavior of OTOC in the gapless region is always similar and does not depend on the anisotropy parameter. We therefore suggest that it is universal in the gapless regime but only at low temperatures. Here we argue that such a universal behavior comes from the fact that the low-energy excitations of the gapless XXZ model are described by the TLL and belong to the same universality class of the 2D CFT with $c=1$~\cite{Korepin,Giamarchi}. We thus expect that the universal decaying behavior of OTOC in the gapless regime should hold for various quantum systems including non-integrable ones if their low-energy spectra are described by the 2D CFT with $c=1$. In fact, the TLL or the 2D CFT with $c=1$ is almost ubiquitous in the low-energy excitations not only for integrable gapless quantum spin chains such as the spin-1/2 XXZ chain but also for non-integrable ones.

Our numerical method is
the infinite time-evolving block decimation (iTEBD) method~\cite{Vidal2003,Vidal2007,Orus2008}, which enables us to directly address the thermodynamic limit. This is a benefit of the iTEBD compared to other methods such as numerical exact diagonalization. We estimate the central charge $c$ by fitting the numerical results with an analytical result of the 2D CFT~\cite{Dubail2017}. As a consequence, we confirmed that $c$ is close to unity, which is consistent with the fact that the XXZ model in the gapless region belongs to the universality class of $c=1$. 

This paper is organized as follows. In Sec.~\ref{averageOTOC}, we explain the averaged OTOC, the OSEE, and the relationship between them. In Sec.~\ref{iTEBD}, we show our main numerical results of the average OTOC in the gapped and the gapless regions in the thermodynamic limit. In Sec.~\ref{sec:summary}, we summarize this paper and make some remarks. In Appendix~\ref{OTOCinequality}, we prove an inequality concerning the averaged OTOC. In Appendix~\ref{ap:caninical_iMPO}, we introduce the canonical form of the infinite matrix product operator (iMPO)~\cite{Verstraete2004,Zwolak2004,Barthel2017}. In Appendix~\ref{ap:numerical}, we discuss the details of numerical errors.

\section{averaged OTOC and OSEE}
\label{averageOTOC}
In this section, we define the averaged OTOC and discuss its relationship to the operator space entanglement entropy (OSEE).

We define the OTOC at finite temperature by Eq.~(\ref{eq:OTOC}).
In the quantum field theory (QFT), the OTOC requires appropriate regularizations and the following regularization is commonly adopted~\cite{Maldacena2016a,Roberts2015}:
\begin{align}
\average{\hat{W}(t)\hat{V}\hat{W}(t)\hat{V}}_\beta \to \tr\qty[\hat{y}\hat{W}(t)\hat{y}\hat{V}\hat{y}\hat{W}(t)\hat{y}\hat{V}],
\end{align}
where $\hat{y}^4 := \mathrm{e}^{-\beta \hat{H}}/Z(\beta)$ and $Z(\beta) := \tr\qty[\mathrm{e}^{-\beta \hat{H}}]$.
We adopt this regularized OTOC, which provides a platform to compare our study and the previous studies of QFT.
 
For simplicity, we assume that the total system is on a lattice and each local site is described by a two-level system (i.e., a qubit). Let S be the set of all sites and $L := |{\rm S}|$ be the number of them. The Hilbert space of the total system is $\mathcal{H}=\qty(\mathbb{C}^2)^{\otimes L}$. 

We consider subsystems $\mathrm{A},\mathrm{B}\subset \mathrm{S}$. The number of sites in A and B are denoted by $a$ and $b$, respectively. These subsystems are defined as $\mathrm{A}:=\{i_1, i_2, \cdots, i_a\}$ and $\mathrm{B}:=\{j_1, j_2, \cdots, j_b\}$. An operator belonging to the Pauli operator basis of A~(B) is respectively written as $\hat{O}_\mathrm{A}$ ($\hat{O}_\mathrm{B}$). The operator is explicitly written by a tensor product $\hat{O}_\mathrm{A} = \hat{o}_{i_1} \otimes \hat{o}_{i_2}\otimes\cdots \otimes\hat{o}_{i_a}$ and $\hat{O}_\mathrm{B} = \hat{o}_{j_1} \otimes \hat{o}_{j_2}\otimes\cdots \otimes\hat{o}_{j_b}$, where $\hat{o}_i \in \qty{ \hat{I}_i,\hat{\sigma}_i^\mathrm{X},\hat{\sigma}_i^\mathrm{Y},\hat{\sigma}_i^\mathrm{Z}} $ is a Pauli matrix at site $i \in {\rm S}$. We denote the sets of the operator basis of A and B as $\mathcal{P}_\mathrm{A}$ and $\mathcal{P}_\mathrm{B}$, respectively.

We now define the averaged OTOC as follows~\cite{Hosur2016,Roberts2017}:
\begin{align}
\label{eq:averagedOTOC}
\overline{F}_{AB}(t) &\coloneqq \frac{1}{2^{2a} \cdot 2^{2b}}\sum_{\hat{O}_\mathrm{A} \in \mathcal{P}_\mathrm{A},\hat{O}_\mathrm{B} \in \mathcal{P}_\mathrm{B}}\tr[\hat{y}\hat{O}_\mathrm{B}(t) \hat{y}\hat{O}_\mathrm{A}(0)\hat{y}\hat{O}_\mathrm{B}(t)\hat{y}\hat{O}_\mathrm{A}(0)],
\end{align}
where we averaged the OTOC over the operator bases $\mathcal{P}_\mathrm{A}$ and $\mathcal{P}_\mathrm{B}$. For the special case of $\beta=0$, it has been shown that the decay of the averaged OTOC is related to the negativity of tripartite mutual information~\cite{Hosur2016}.

We introduce a basic property of the averaged OTOC. For A and B, we consider subsystems $\mathrm{A}'\subseteq \mathrm{A}$ and $\mathrm{B}'\subseteq \mathrm{B}$, respectively. Then, the averaged OTOC satisfies
\begin{align}
\label{eq:OTOCinequality}
\overline{F}_{\mathrm{AB}}(t)\leq\overline{F}_{\mathrm{A}'\mathrm{B}'}(t)
\end{align}
(see the Appendix~\ref{OTOCinequality} for the proof).
This inequality holds even when A and B have an intersection.

In the same setup, we define the OSEE. For the Hilbert space of the total system $\mathcal{H}$, we consider a copy of $\mathcal{H}$ and denote it as $\mathcal{\tilde{H}}$. We regard a linear operator $\hat{W}$ acting on $\mathcal{H}$ as a state on the extended Hilbert space $\mathcal{H}\otimes \mathcal{\tilde{H}}$ as 
\begin{align}
\label{Operator state}
\ket{\hat{W}} \coloneqq  \hat{W}\otimes \hat{I} \ket{\Phi},
\end{align}
where $\ket{\Phi}\coloneqq \prod_{i=1}^{L} (\ket{0}_i\ket{0}_{\tilde{i}}+\ket{1}_i\ket{1}_{\tilde{i}})$ is the EPR pair between the original system and the copied system and $\hat{I}$ is the identity operator. Here, the index $i$ represents the original system and $\tilde{i}$ represents its copy. We note that this mapping is essentially the same as the channel-state duality~\cite{Choi}. When the total system is split into A and B, i.e., $\mathrm{S}=\mathrm{A}\cup \mathrm{B}, \mathrm{A}\cap \mathrm{B} =\emptyset$, the OSEE of the operator $\hat{W}$ in the region A is defined as
\begin{align}
{S_\mathrm{A}^{\mathrm{OSEE}}}\qty[\hat{W}] &\coloneqq-\tr_{\mathrm{A}\tilde{\mathrm{A}}}\qty[\hat{\rho} \ln\hat{\rho}],
\end{align}
where
\begin{align}
\hat{\rho} &\coloneqq\frac{ \tr_{\mathrm{B}\tilde{\mathrm{B}}}\qty[\ket{\hat{W}}\bra{\hat{W}}]}{\braket{\hat{W}}{\hat{W}}},
\end{align}
with $\tilde{\mathrm{A}}$ being the copy of $\mathrm{A}$ and $\tilde{\mathrm{B}}$ being the copy of $\mathrm{B}$.
Furthermore, the R\'enyi-$n$ OSEE of $\hat{W}$ is defined as
\begin{align}
{S_\mathrm{A}^{\mathrm{OSEE}\:(n)}}\qty[\hat{W}] \coloneqq \frac{1}{1-n}\log( \frac{\tr\left[ \left(\hat{W}^\dagger\right)^{\otimes n}\hat{S}_\mathrm{A}^{(1,n,n-1,\cdots,2)} \hat{W}^{\otimes n } \hat{S}_{\mathrm{A}}^{(1,2,3,\cdots,n)}\right]}{\tr\left[\hat{W}^\dagger \hat{W}\right]}),
\end{align}
where $\hat{S}_\mathrm{A}^{(1,2,3,\cdots,n)}$ is a permutation operator acting on $\mathcal{H}^{\otimes n}$, which cyclically permutes the replicas of the subsystem A. Explicitly, the action of $\hat{S}_\mathrm{A}^{(1,2,3,\cdots,n)}$ is given by $\hat{S}_\mathrm{A}^{(1,2,3,\cdots,n)}\ket{i_1}_\mathrm{A} \otimes\ket{j_1}_\mathrm{B} \otimes\ket{i_2}_\mathrm{A}\otimes\ket{j_2}_\mathrm{B} \otimes \cdots \otimes \ket{i_n}_\mathrm{A} \otimes \ket{j_n}_\mathrm{B} =\ket{i_2}_\mathrm{A}\otimes \ket{j_1}_\mathrm{B} \otimes \ket{i_3}_\mathrm{A}\otimes\ket{j_2}_\mathrm{B} \cdots \otimes \ket{i_n}_\mathrm{A} \otimes \ket{j_{n-1}} \otimes\ket{i_1}_\mathrm{A} \otimes \ket{j_n}_\mathrm{B}$, where $\ket{i}_\mathrm{A}$ belongs to a basis of the subsystem A and $\ket{j}_\mathrm{B}$ belongs to a basis of the subsystem B. We note that $\hat{S}_\mathrm{A}^{(1,n,n-1,\cdots,2)}$ is the inverse operator of $\hat{S}_\mathrm{A}^{(1,2,3,\cdots,n)}$.

The averaged OTOC has the following relation to the OSEE~\cite{Hosur2016}.
When $\mathrm{S}=\mathrm{A}\cup \mathrm{B}$ and $\mathrm{A}\cap \mathrm{B} =\emptyset $,
the averaged OTOC is written as
\begin{align}
\label{eq:OTOCtoOSEE}
\overline{F}_{\mathrm{A}\mathrm{B}}(t) &=\frac{Z(\beta/2)^2}{2^LZ(\beta)}\exp\qty(-{S_\mathrm{A}^{\mathrm{OSEE\:(2)}}}\qty[\hat{U}(\beta,t)] ),
\end{align}
where $\hat{U}(\beta,t)\coloneqq \mathrm{e}^{-\left(\frac{\beta}{4}+\mathrm{i}t\right)\hat{H}}$.
From Eq.~(\ref{eq:OTOCinequality}), 
\begin{align}
\frac{Z(\beta/2)^2}{2^LZ(\beta)}\mathrm{e}^{-S_\mathrm{A}^{\mathrm{OSEE\:(2)}}}\leq\overline{F}_{\mathrm{A}'\mathrm{B}'}(t)
\end{align}
holds for $\mathrm{A}'\subseteq \mathrm{A}$ and $\mathrm{B}'\subseteq \mathrm{B}$.
Thus, the averaged OTOC $\overline{F}_{\mathrm{A}'\mathrm{B}'}(t)$ is bounded from below with the R\'enyi-2 OSEE. Although the OTOC with local observables is investigated in many previous studies, we focus on Eq.~(\ref{eq:OTOCtoOSEE}) to consider a lower bound of the averaged OTOC.

The OSEE is also investigated in QFT. The OSEE of $\hat{U}(\beta,t)$ in quantum critical systems has been calculated analytically by the 2D CFT~\cite{Dubail2017}. In QFT, the total system is not on a lattice and subsystems are represented by intervals. With the bipartition $\mathrm{A} = \left(-\infty,0\right]$ and $\mathrm{B} = \left(0,\infty \right)$, the OSEE of $\hat{U}(\beta,t)$ is given by
\begin{align}
\label{eq:OSEE}
{S_\mathrm{A}^{\mathrm{OSEE\:(n)}}}\qty[\hat{U}(\beta,t)] =\frac{c}{6}\qty(1+\frac{1}{n})\ln(\cosh(\frac{2\pi t}{\beta}))+S_0,
\end{align}
where $S_0$ is an unimportant constant. This formula is derived with the replica trick as the case of the Calabrese-Cardy formula~\cite{Calabrese2004}. We make a remark on the case of another partition $\mathrm{A} = \left[0,L_\mathrm{A}\right]$ and  $\mathrm{B}=\mathrm{S}\setminus \mathrm{A}$. In the limit of $L_\mathrm{A} \gg 1$, the OSEE ${S_\mathrm{A}^{\mathrm{OSEE}\:(n)}}\qty[\hat{U}(\beta,t)]$ becomes double of Eq.~(\ref{eq:OSEE}) because there are two boundaries. 

We obtain an expression of the averaged OTOC in terms of the OSEE by substituting Eq.~(\ref{eq:OSEE}) into Eq.~(\ref{eq:OTOCtoOSEE}):
\begin{align}
\label{eq:OSEEOTOC}
\overline{F}_{\mathrm{AB}}(t) \propto \exp\qty(-{S_\mathrm{A}^{\mathrm{OSEE\:(2)}}}\qty[\hat{U}(\beta,t)] )\propto\frac{1}{\qty(\cosh(\frac{2\pi t}{\beta }))^{\frac{c}{4}}}.
\end{align}
In the same manner, for the partition with $\mathrm{A} = \left[0,L_\mathrm{A}\right]$, the averaged OTOC is obtained as $\overline{F}_{\mathrm{AB}}(t) \propto \qty(\cosh(\frac{2\pi t}{\beta }))^{-c/2}$. 

At early times $t/\beta\ll 1$, the above expression of the averaged OTOC is different from $F(t)\approx a-b\mathrm{e}^{\lambda_L t}$ that is expected from the chaotic bound~\cite{Maldacena2016,Kitaevtalk}. This discrepancy comes from the fact that we did not take the semi-classical limit (e.g., the large-$c$ limit). 

On the other hand, at late times $t/\beta\gg 1$, Eq.~(\ref{eq:OSEEOTOC}) reduces to the exponential decay $\overline{F}_{AB}(t)\sim\mathrm{e}^{-\frac{c \pi t}{ \beta}}$. We note that from the 2D CFT with the large-$c$ limit~\cite{Roberts2015}, the OTOC with the primary fields $\hat{V}$ and $\hat{W}$ also decays exponentially at $t/\beta\gg 1$: 
\begin{align}
\label{eq:OTOC_CFT_LATE_TIME}
\frac{\average{\hat{W}(t)\hat{V}\hat{W}(t)\hat{V}}_\beta}{\average{\hat{W}\hat{W}}_\beta\average{\hat{V}\hat{V}}_\beta} \sim \mathrm{e}^{-4\pi h_v t},
\end{align}
where $h_v$ is the conformal dimension of $V$. 

The exponent in Eq.~(\ref{eq:OTOC_CFT_LATE_TIME}) is different from that of the averaged OTOC, because the conformal dimension of the twist operator for $n=2$ is $c/16$. This is not a contradiction and our results are consistent with Eq.~(\ref{eq:OTOC_CFT_LATE_TIME}). This is because Eq.~(\ref{eq:OTOC_CFT_LATE_TIME}) is obtained in the parameter region $1\ll h_v\ll h_w$~\cite{Roberts2015} with $h_w$ being the conformal dimension of $W$, which is not satisfied in our setup. Besides, the positions of the primary fields in the spacetime in Ref.~\cite{Roberts2015} are different from those in our setup. 

\section{Numerical results by iTEBD}
\label{iTEBD}
In this section, we present our main numerical results.  Specifically, by using the iTEBD, we calculate the OSEE of $\hat{U}(\beta,t)=\mathrm{e}^{-\left(\frac{\beta}{4}+\mathrm{i}t\right)\hat{H}}$ and discuss the qualitative difference between the gapless and gapped regions.

\subsection{Setup}
In this subsection, we describe our setup for the numerical calculation and the protocol of the iTEBD. The Hamiltonian of the one-dimensional quantum spin-$1/2$ anti-ferromagnetic XXZ chain is defined as
\begin{align}
\hat{H}_{\mathrm{XXZ}}
:=
J\sum_{i=-\infty}^\infty\qty[\hat{\sigma}_i^\mathrm{X}\hat{\sigma}_{i+1}^\mathrm{X}+\hat{\sigma}_i^\mathrm{Y}\hat{\sigma}^\mathrm{Y}_{i+1}+\Delta\qty(\hat{\sigma}_i^\mathrm{Z}\hat{\sigma}_{i+1}^\mathrm{Z}-1)],
\end{align}
where $\hat{\sigma}_i^\mathrm{X},\hat{\sigma}_i^\mathrm{Y}$ and $\hat{\sigma}_i^\mathrm{Z}$ are the Pauli operators at site $i$, $\Delta$ is the anisotropy parameter, and $J$~$(>0)$ is the strength of the exchange interaction. For simplicity, we set $J=1$. For $0\leq \Delta \leq 1$, the low energy excitations from the ground state are gapless and this model exhibits quantum criticality. From the 2D CFT, it is known that the gapless XXZ model belongs to the universality class with the conformal charge $c=1$~\cite{Giamarchi,Korepin}. We divide the total system into two semi-infinite subsystems $\mathrm{A}=\qty{-\infty,\cdots,-1,0}$ and $\mathrm{B}=\qty{1,2,\cdots\infty}$ and calculate the R\'enyi-2 OSEE for this bipartition.

\begin{figure}[th]
	\centering
	\includegraphics[width=1.0\linewidth]{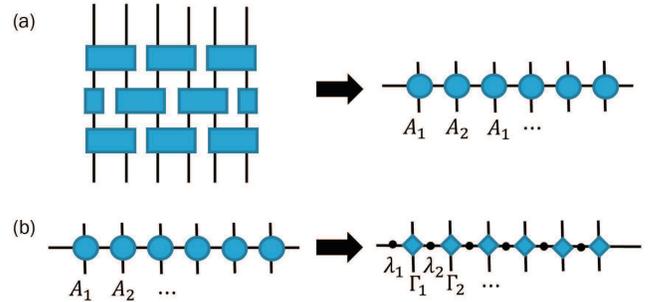}
	\caption{Schematics of the iTEBD. (a) After the Trotter decomposition of $\mathrm{e}^{-\frac{\beta}{4}\hat{H}}$, we obtain the two-site translation invariant iMPO composed of two tensors $\qty{A_1,A_2}$ by performing the SVDs. (b) Transformation the iMPO into the canonical form composed of four tensors.}
	\label{fig:MPOaprrox}
\end{figure}

We explain the protocol to approximately calculate the imaginary-time evolution operator by the iTEBD. We perform the Trotter decomposition~\cite{Trotter1959} of $\mathrm{e}^{-\frac{\beta}{4}\hat{H}}$ and obtain the products of $\mathrm{e}^{-\Delta \beta \hat{H}}$, where $\Delta\beta:=\beta/(4N_\mathrm{Trotter})$ and $N_\mathrm{Trotter}$ is the number of decomposition. In this study, we adopt the Trotter decomposition of the second order
\begin{align}
\mathrm{e}^{h(\hat{A}+\hat{B})} =\mathrm{e}^{h\hat{A}/2} \mathrm{e}^{h\hat{B}}\mathrm{e}^{h\hat{A}/2}+ \order{h^3},
\end{align}
where we set $h=-\Delta\beta$. We split $\hat{H}=\hat{A}+\hat{B}$ into $\hat{A}$ and $\hat{B}$. Then, $\hat{A}$ ($\hat{B}$) corresponds to the interactions defined in the odd (even) bonds, respectively.

Then, we approximate the products of the infinitesimal time evolution operators in the form of the iMPO by repeatedly using the singular value decomposition (SVD). In general, the MPO provides a representation of a many-body operator, which decomposes it into a one-dimensional sequence of locally defined tensors. With the bond dimension $\chi$, we control the entanglement in the operator space. By assuming the translational invariance, the MPO can be applied to an infinite chain and the extended MPO for that case is referred to as the iMPO. Figure~\ref{fig:MPOaprrox}(a) shows the approximation protocol of $\mathrm{e}^{-\frac{\beta}{4}\hat{H}}$ by the iTEBD and the iMPO. By construction of the iTEBD protocol with the Trotter decomposition, we finally obtain a translational invariant iMPO composed of two tensors $\qty{A_1,A_2}$ with the two-site unit cell.

After performing the iTEBD,  we transform the iMPO into the canonical form~\cite{Orus2008} to calculate the OSEE. As illustrated in Fig.~\ref{fig:MPOaprrox}(b), the canonical form for the two-site translational invariant case is composed of four tensors $\qty{\lambda_{\mathrm{A}_1},\Gamma_{\mathrm{A}_1},\lambda_{\mathrm{A}_2},\Gamma_{\mathrm{A}_2}}$,
where $\Gamma_{\mathrm{A}_1}$ and $\Gamma_{\mathrm{A}_2}$ are complex-valued tensors with four indices, and $\lambda_{\mathrm{A}_1}$ and $\lambda_{\mathrm{A}_2}$ are diagonal matrices with non-negative elements. The definition of the canonical form is explained in Appendix~\ref{ap:caninical_iMPO}. Using the canonical form, we can calculate the R\'enyi-2 OSEE for the bipartition $\mathrm{A}=\qty{-\infty,\cdots,-1,0}$ and $\mathrm{B}=\qty{1,2,\cdots\infty}$.

We can also approximate the complex-time evolution operator $\hat{U}(\beta,t)$ in the same manner. In the case of the complex-time evolution, we first approximate $\hat{U}(\beta,0)$ in the form of the iMPO. By taking it as the initial state, we execute the iTEBD for $\hat{U}(0,t)$.

We note that the MPO formalism itself does not work very efficient for the direct calculation of the OTOC~\cite{Hemery2019}. In our study, we calculated OSEE by using Eq.~(\ref{eq:OSEEOTOC}) which relates the OSEE and the averaged OTOC.

\begin{figure}[th]
	\centering
	\includegraphics[width=1.0\linewidth]{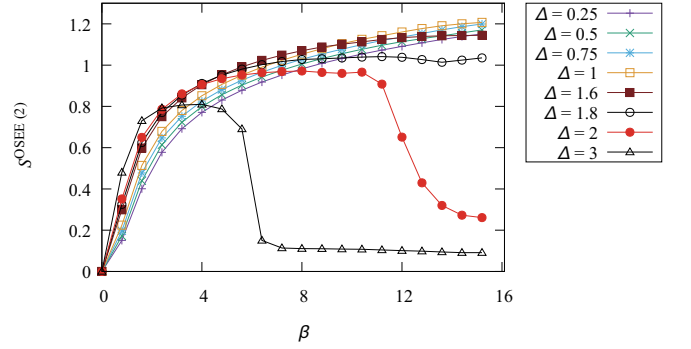}
	\caption{
	Imaginary time evolution of the OSEE ${S_A^{\mathrm{OSEE\:(2)}}}$ for the operator $\mathrm{e}^{-\frac{\beta}{4}\hat{H}}$ obtained by the iTEBD. The bond dimension $\chi$ is $512$ and the Trotter decomposition width is $\Delta \beta=0.01$. Small anisotropy parameters $\Delta=0.25,0.5,0.75,1$ correspond to the gapless region and large ones $\Delta=1.6,1.8,2,3$ correspond to the gapped region. As the temperature decreases, the difference between these regions becomes more visible.}
		\label{fig:temperatureOSEE}
\end{figure}

\subsection{Imaginary-time evolution}
\label{sec:imagtime}

We now show our numerical results. Figure~\ref{fig:temperatureOSEE} shows the imaginary-time dependence of the R\'enyi-2 OSEE ${S_A^{\mathrm{OSEE\:(2)}}}$ for $\hat{U}(\beta,0)=\mathrm{e}^{-\frac{\beta}{4}\hat{H}}$. The anisotropy parameters are taken as $\Delta=0.25, 0.5, 0.75, 1, 1.6,1.8, 2, 3$. We note that a similar calculation of imaginary time evolution has been performed for a finite chain~\cite{Barthel2017}, while our results are for an infinite chain in the thermodynamic limit.

In the high-temperature region with small $\beta$, the OSEE increases monotonically as $\beta$ increases regardless of the existence of the gap. However, in the low-temperature region with large $\beta$, the difference between the gapless and gapped regions becomes visible. While the OSEE continues to increase monotonically in the gapless region, the OSEE in the gapped region with $1< \Delta$ shows non-monotonic behavior. This result implies that the OSEE captures the difference between the excitation structure near the ground state, which is accessible if the temperature is sufficiently low.

We discuss the reason why the OSEE in the gapped region exhibits the non-monotonic behavior. The XXZ model in the gapped region has the two-fold degeneracy in the ground states, which exhibits the $\mathrm{Z}_2$ symmetry. Let $E_0$ be the eigenenergy of the ground states and $\ket{\psi^0_{\mathrm{GS}}}$ and $\ket{\psi^1_{\mathrm{GS}}}$ be the corresponding eigenvectors. In the gapped region and at sufficiently low temperature $\beta \gg 1$, we obtain $\mathrm{e}^{-\beta \hat{H}}\simeq \mathrm{e}^{-\beta E_0}\qty(\ket{\psi^0_{\mathrm{GS}}} \bra{\psi^0_{\mathrm{GS}}}+ \ket{\psi^1_{\mathrm{GS}}} \bra{\psi^1_{\mathrm{GS}}})$. Thus, the corresponding state vector in the extended Hilbert space is approximated as
\begin{align}
\ket{\mathrm{e}^{-\beta \hat{H}}} \simeq \frac{1}{\sqrt{2}}\qty(\ket{\psi^0_{\mathrm{GS}}} \otimes\ket{\psi^0_{\mathrm{GS}}}+ \ket{\psi^1_{\mathrm{GS}}} \otimes\ket{\psi^1_{\mathrm{GS}}}).
\end{align}

In our numerical calculation, however, the iMPO obtained by the iTEBD is two-site translation invariant, while the operator $\mathrm{e}^{-\frac{\beta}{4}\hat{H}}$ is one-site translation invariant. This implies that the one-site translation invariance is not guaranteed and can be broken in our results. In such a case, the ground-state degeneracy is removed by the artificial symmetry breaking, resulting in an artificial non-monotonic behavior in numerical calculations~\cite{Jiang2013}. 

Due to the removal of the degeneracy, $\ket{\mathrm{e}^{-\beta \hat{H}}}$ is described by only one of $\ket{\psi^0_{\mathrm{GS}}}$ or $\ket{\psi^1_{\mathrm{GS}}}$. This is the reason why the OSEE in the gapped region shows the non-monotonic $\beta$-dependence in Fig.~\ref{fig:temperatureOSEE}. As the temperature decreases, the removal of the degeneracy becomes more visible, and the OSEE decays more rapidly.
As a special case, if the anisotropy parameter is very large $\Delta \gg 1$, the degenerate ground states are represented as $\ket{\psi^0_{\mathrm{GS}}}\simeq\ket{\uparrow \downarrow \uparrow \downarrow \cdots}$ and $\ket{\psi^1_{\mathrm{GS}}}\simeq\ket{\downarrow \uparrow \downarrow \uparrow\cdots}$. In this case, the OSEE of $\mathrm{e}^{-\beta \hat{H}}$ at $\beta \gg 1$ is given by
\begin{align}
\label{eq:OSEElog2}
\begin{aligned}
&{S_A^{\mathrm{OSEE}\:(n)}}\qty[\mathrm{e}^{-\frac{\beta}{4} \hat{H}}] \\
&\simeq\frac{1}{2}{S_A^{\mathrm{OSEE}\:(n)}}\qty[\ket{\psi^0_{\mathrm{GS}}}\bra{\psi^0_{\mathrm{GS}}}] + \frac{1}{2}{S_A^{\mathrm{OSEE}\:(n)}}\qty[\ket{\psi^1_{\mathrm{GS}}}\bra{\psi^1_{\mathrm{GS}}}]+\ln 2 \\
& ={S_A^{\mathrm{OSEE}\:(n)}}\qty[\ket{\psi^0_{\mathrm{GS}}}\bra{\psi^0_{\mathrm{GS}}}]+\ln 2,
\end{aligned}
\end{align}
where we used the fact that the entanglement entropies of the two ground states are the same to obtain the third line. When the degeneracy is removed, the OSEE decreases by $\ln 2$ in the above equation, because the OSEE reduces to that of a single ground state. 

\begin{figure}
	\centering
	\includegraphics[width=1\linewidth]{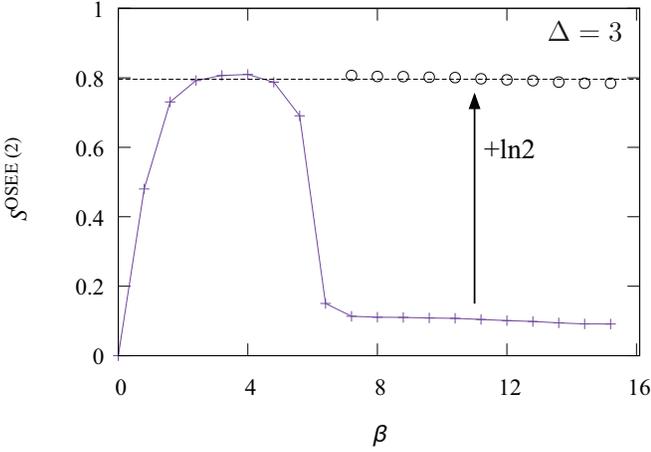}
	\caption{OSEE ${S_A^{\mathrm{OSEE\:(2)}}}$ for $\mathrm{e}^{-\frac{\beta}{4}\hat{H}}$ with $\Delta=3$ in Fig.~\ref{fig:temperatureOSEE} is extracted. The open black circles represents ${S_A^{\mathrm{OSEE\:(2)}}}+\ln 2$ $(7.2 \le \beta)$ and the black dashed line represents the average of them. If there is no artificial symmetry breaking, the OSEE at low temperature would saturate near the black dashed line.}
	\label{fig:temperatureOSEE_partial}
\end{figure}

Figure~\ref{fig:temperatureOSEE_partial} shows the inverse-temperature dependence of the OSEE with $\Delta=3$, which is already shown in Fig.~\ref{fig:temperatureOSEE}. The open black circles depict ${S_A^{\mathrm{OSEE\:(2)}}}+\ln 2$ at low temperature $(7.2 \le \beta)$ and the black dashed line shows the average of them. We see that if Eq.~(\ref{eq:OSEElog2}) holds approximately, the OSEE at low temperature saturates near the black dashed line.

\subsection{Real-time evolution}
\label{sec:realtime}
We next consider the real-time evolution of the OSEE.  In the following, we adopt sufficiently high temperatures where the above-mentioned artificial symmetry breaking does not occur.

\begin{figure*}
	\centering
	\subfigure{\includegraphics[width=0.4\linewidth]{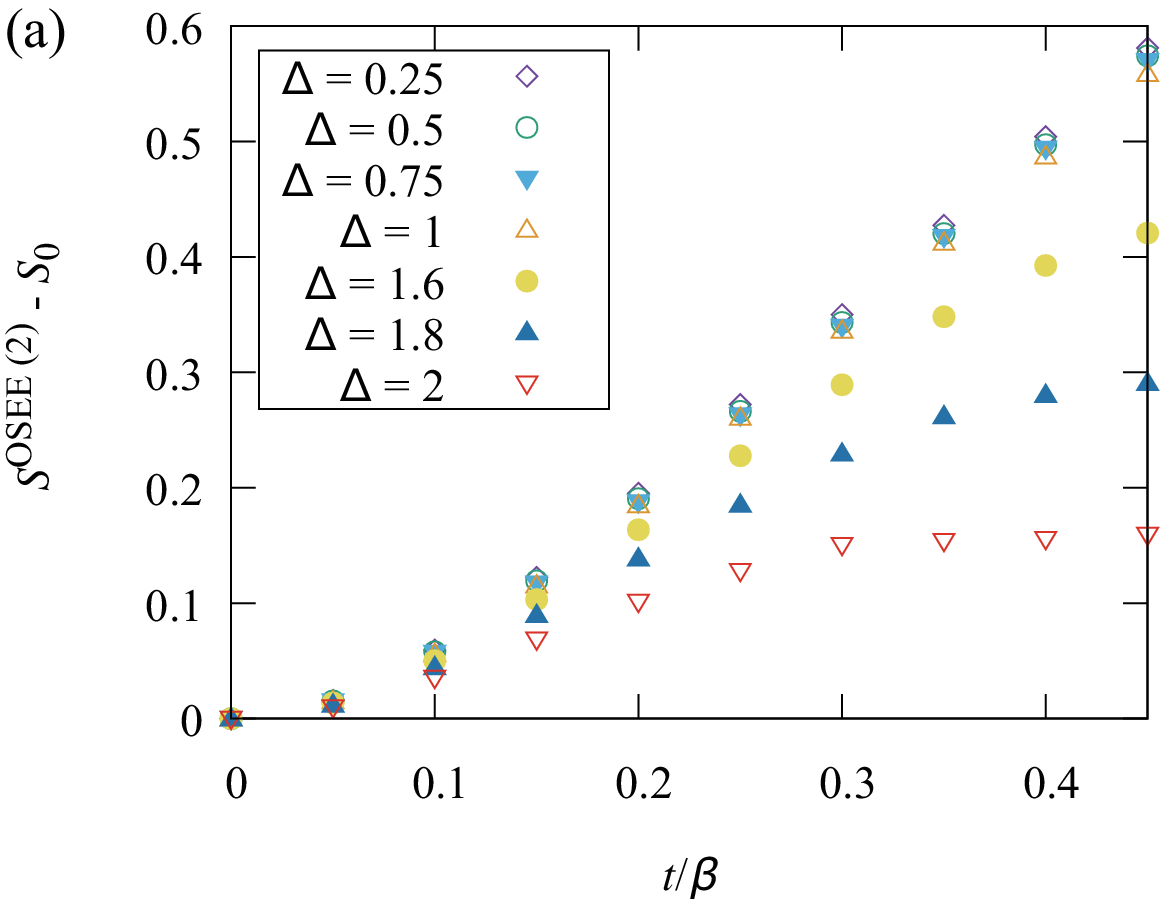}
		\label{fig:timeOSEE}}
	\hspace{20mm}
	\subfigure{\includegraphics[width=0.4\linewidth]{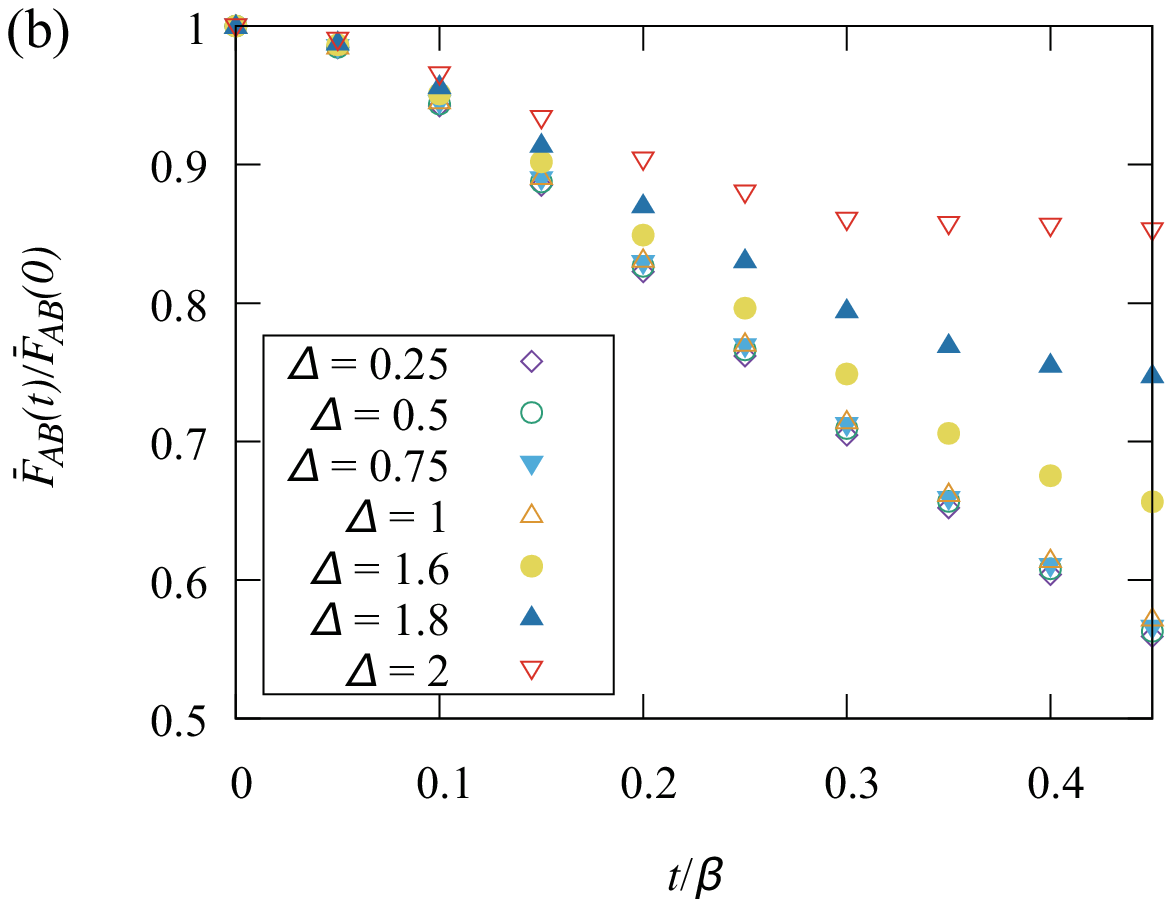}
		\label{fig:timeOTOC}}
	\caption{(a) Real-time dependence of the OSEE ${S_\mathrm{A}^{\mathrm{OSEE\:(2)}}}$ for the complex-time evolution operator $\mathrm{e}^{-\left(\frac{\beta}{4}+\mathrm{i}t\right)\hat{H}}$ for the XXZ model. We subtracted $S_0$, which is the value of the OSEE at $t=0$. (b) Real time dependence of the averaged OTOC  $\overline{F}_{\mathrm{AB}}$ corresponding to the OSEE by Eq.~(\ref{eq:OTOCtoOSEE}). The vertical axis is normalized by dividing by $\overline{F}_{\mathrm{AB}}(0)$. The bond dimension $\chi$ is $512$ and the Trotter decomposition widths on the imaginary time axis and the real time axis are $\Delta \beta=\Delta t=0.005$. Small anisotropy parameters $\Delta=0.25,0.5,0.75,1$ correspond to the gapless region and large ones $\Delta=1.6,1.8,2$ correspond to the gapped region. The averaged OTOC $\overline{F}_{\mathrm{AB}}$ decays faster in the gapless region. The averaged OTOCs in the gapless region decay in the same manner.}
	\label{fig:timeOTOCandOTOC}
\end{figure*}

Figure~\ref{fig:timeOTOCandOTOC} (a) shows the real-time dependence of the OSEE for $\mathrm{e}^{-\left(\frac{\beta}{4}+\mathrm{i}t\right)\hat{H}}$ at $\beta=4$ and with $\Delta=0.25, 0.5, 0.75, 1, 1.6, 1.8, 2$. The OSEE in the gapped region increases less than that in the gapless region. This implies that the corresponding averaged OTOC decays faster in the gapless region. 

In the gapless region ($\Delta=0.25, 0.5, 0.75, 1$), the averaged OTOC exhibits the same behavior for the different values of $\Delta$. As we will argue in the next subsection, this is because the XXZ model in the gapless region belongs to the same universality class with the central charge $c=1$. In the gapped region, the energy gap  $E_\mathrm{G}$ is written as an increasing function of $\Delta$: $E_\mathrm{G}\simeq 4\pi J\exp \qty{-\frac{\pi^2}{2\sqrt{2\qty(\Delta -1)}}}$~\cite{Baxter,Takahashi}. From Fig.~\ref{fig:timeOTOCandOTOC} (a), the increase of the OSEE is smaller if $E_\mathrm{G}$ is larger. This implies that the decay of the averaged OTOC becomes slower, as the parameter goes apart from the gapless region and the energy gap $E_\mathrm{G}$ becomes larger. 

\subsection{Role of the scale invariance}
\label{sec:scale_invariance}

The numerical result in previous subsections implies that the system with the scale invariance scrambles fast. We will further investigate the role of scale invariance. 

In the context of the 2D CFT, the R\'enyi-$n$ OSEE is represented by a correlation function of the twist operators on a cylinder. Let $L_\mathrm{eff}=\frac{\beta v_s}{2\pi}\cosh\frac{2 \pi}{\beta}t$ be the effective length. The two-point function of the twist operator is given by
\begin{align}
\average{\mathcal{T}(u)\overline{\mathcal{T}}(v)}_{\mathrm{cyl}} \propto 	L_{\mathrm{eff} }^{-\Delta_n},
\end{align}
where $\Delta_n=\frac{c}{12}(n-\frac{1}{n})$. The OSEE with the bipartition $\mathrm{A} = \left(-\infty,0\right]$ and $\mathrm{B} = \left(0,-\infty \right)$ is written as ${S_A^{\mathrm{OSEE}\:(n)}}\qty[\mathrm{e}^{-\frac{\beta}{4} \hat{H}}]=\frac{1}{1-n}\ln \average{\mathcal{T}(u)\overline{\mathcal{T}}(v)}_{\mathrm{cyl}}$, and the R\'enyi-$n$ OSEE is given by ${S_A^{\mathrm{OSEE\:(2)}}}=\frac{c}{2}\ln L_{\mathrm{eff}}+a$, where $a$ is a constant. Thus, we expect that the numerical results (even with different temperatures) of the OSEE are characterized by a single function that depends only on $L_{\mathrm{eff}}$ if the scale invariance emerges.

\begin{figure}[th]
	\centering
	\includegraphics[width=1\linewidth]{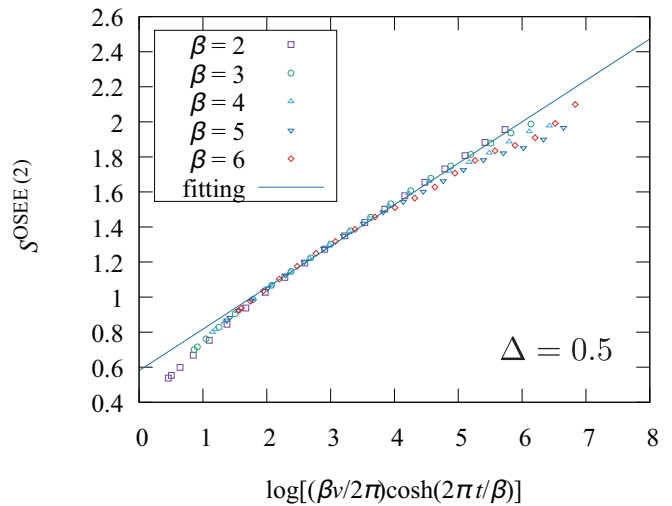}
	\caption{
	The effective length dependence of the R\'enyi-2 OSEE ${S_A^{\mathrm{OSEE\:(2)}}}$ for the complex-time evolution operator $\mathrm{e}^{-\left(\frac{\beta}{4}+\mathrm{i}t\right)\hat{H}}$ in the gapless region ($\Delta=0.5$). The horizontal axis is $\ln L_{\mathrm{eff}}=\ln\qty(\frac{\beta v}{\pi} \cosh(\frac{2\pi}{\beta}t))$. For each of the inverse temperatures $\beta=2,3,4,5,6$, we numerically calculate the real-time evolution by fixing $\beta$. The solid line is obtained by the fitting with the fitting range $[3,4]$.
}
	\label{fig:timeOSEE_Jz0.5}
\end{figure}

Figure~\ref{fig:timeOSEE_Jz0.5} shows the effective-length dependence of the OSEE for the complex-time evolution operator $\mathrm{e}^{-\left(\frac{\beta}{4}+\mathrm{i}t\right)\hat{H}}$ in the gapless region ($\Delta=0.5$). The horizontal axis is $\ln L_{\mathrm{eff}}=\ln\qty(\frac{\beta v_s}{2\pi} \cosh(\frac{2\pi}{\beta}t))$. We plot the R\'enyi-2 OSEE at $\beta=2,3,4,5,6$. For each parameter, we calculate the effective length dependence by fixing the inverse temperature and by changing the real time. As shown in Fig.~\ref{fig:timeOSEE_Jz0.5}, the data points almost collapse onto a single curve, which implies that the OSEE depends only on the effective length $L_\mathrm{eff}$. In the right side of the figure, however, we find that there are some deviations. We consider that these deviations are caused by the fact that the bond dimension $\chi$ used in our numerical calculation is not large enough. In Appendix~\ref{ap:numerical}, we evaluate the numerical errors from the Trotter decomposition and the bond dimension, where the numerical errors are sufficiently small in the range of $\ln L_{\mathrm{eff}} \le 4$. 

We fit the numerical data points for $\beta=2,3,4,5,6$ against the fitting function ${S_A^{\mathrm{OSEE\:(2)}}}=\frac{c}{4}\ln L_{\mathrm{eff}}+a$, which is from the 2D CFT with fitting parameters $c$ and $a$. The fitting range is taken as $\ln L_{\mathrm{eff}} \le 4$. We expect that the universal behavior of the correlation function appears when $L_{\mathrm{eff}}$ is sufficiently large. 

\begin{table}[thb]
	\label{tab:fitting}
	\centering
	\caption{Fitting results for the OSEE shown in Fig.~\ref{fig:timeOSEE_Jz0.5} with a fitting function ${S_A^{\mathrm{OSEE\:(2)}}}=\frac{c}{4}\ln L_{\mathrm{eff}}+a$. 
	}
	\begin{tabular}{|l||c|c|}
		Fitting range & $c$ &  $a$ \\  \hline
		$[2.5,4]$& 0.957 $\pm$0.008 &0.571 $\pm$ 0.007   \\
		$[3,4]$& 0.946 $\pm$0.016 &0.581 $\pm$ 0.014   \\
		$[3.5,4]$& 0.96 $\pm$0.06 &0.569 $\pm$ 0.05   
		\label{tab:c_a}
	\end{tabular}
\end{table}

We perform the fitting by changing the lower end of the fitting range within $\ln L_{\mathrm{eff}} \le 4$, i.e., for intervals $[2.5,4]$, $[3,4]$, and $[3.5,4]$. The results are summarized in Table~\ref{tab:c_a}. The obtained values of $c$ are close to unity, which is consistent with the fact that the XXZ model with $\Delta=0.5$ at low temperature is a quantum critical system described by the 2D CFT with the central charge $c=1$. Thus, we conclude that the numerical results in the gapless region show universal behavior caused by the scale invariance.

\section{Summary and discussion}
\label{sec:summary}
We have investigated the relationship between the scale invariance at quantum criticality and the OTOC for the one-dimensional anti-ferromagnetic quantum spin-$1/2$ XXZ chain. Specifically, we have numerically calculated the time evolution of the OSEE for $\hat{U}(\beta,t)=\mathrm{e}^{-\left(\frac{\beta}{4}+\mathrm{i}t\right)\hat{H}}$ both in the gapless region and in the gapped region. 

We have shown that the OSEE exhibits qualitatively different behaviors in the gapless and the gapped regions as demonstrated in Fig.~\ref{fig:timeOTOCandOTOC}. In the gapped region, the amount of increase of the OSEE becomes smaller as the system goes away from the gapless region, which implies that the decay of the averaged OTOC becomes slower. 
While these numerical results are for bipartition of the chain (i.e., $\mathrm{A}=\qty{-\infty,\cdots,-1,0}$ and $\mathrm{B}=\qty{1,2,\cdots\infty}$), we expect a similar behavior for general $\mathrm{A}^\prime$, $\mathrm{B}^\prime$ because of inequality Eq.~(\ref{eq:OTOCinequality}). Furthermore, as shown in Fig.~\ref{fig:timeOSEE_Jz0.5}, the OSEE for $\hat{U}(\beta,t)$ in the gapless region at low temperature exhibits the similar behavior as that derived by the 2D CFT analytically, which is consistent with the fact that the 2D CFT is an effective theory of the XXZ model.

In this study, we have performed numerical simulations in the range of $\Delta\geq 0$. We argue that the averaged OTOC in $-1<\Delta<0$ behaves similarly to that in $0\leq\Delta\leq 1$, because the low-energy physics is effectively described by the 2D CFT of $c=1$ in all of the gapless regime $-1<\Delta\leq 1$~\cite{Korepin,Giamarchi}. In $\Delta\leq -1$, on the other hand, it is known that the low-energy excitation above the ground state is gapped, where magnon bound states appear~\cite{Fukuhara2013}. Since the propagation velocity of free magnons is faster than that of magnon bound states, the decay of the OTOC in the early time regime with $\Delta < -1$ is dominated by the contribution of free magnons and thus similar to that in the gapped region $\Delta>1$.

On the basis of our numerical results, we have concluded that the scale invariance (or quantum criticality) is important for the fast decay of the OTOC. On the other hand, we did not observe that the OTOC in the XXZ model exhibits the exponential decay and saturates the MSS bound~\cite{Maldacena2016}, in contrast to the SYK model~\cite{Maldacena2016a} and the 2D CFT~\cite{Roberts2015}. This is consistent with previous results~\cite{Maldacena2016}, because the saturation of the MSS bound would come from the semiclassical limit such as the large central charge limit $c\gg 1$ of the 2D CFT, while this limit cannot be taken in our setup of the XXZ spin chain. Investigating whether the MSS bound can be saturated in a spin model with local interactions is a future issue.
The quantum simulator can experimentally address this problem with ultracold atoms~\cite{Gross2017} and superconducting qubits~\cite{Wendin2017}. 

\begin{acknowledgments}
We are grateful to Shunsuke Furukawa, Youhei Yamaji, and Masatoshi Imada for valuable discussions.
E.I. and T.S. are supported by JSPS KAKENHI Grant Number JP16H02211.
E.I. is also supported by JSPS KAKENHI Grant Number JP15K20944.
\end{acknowledgments}
\appendix
\section{Inequality for OTOC}
\label{OTOCinequality}
In this Appendix, we prove inequality~(\ref{eq:OTOCinequality}). 
We first discuss the representation of the OTOC in the operator space~\cite{Xu}. By using the notation of Eq.~(\ref{Operator state}), we write the inner product between operators $\hat{O}_\mathrm{A}$ and $\hat{O}_\mathrm{B}$ as
\begin{align}
\braket{\hat{O}_\mathrm{A}}{\hat{O}_\mathrm{B}} = \tr\left[\hat{O}_\mathrm{A}^\dagger \hat{O}_\mathrm{B}\right].
\end{align}
Also, the product $\hat{O}_\mathrm{A}\hat{O}_\mathrm{B}$ can be written as the following state:
\begin{align}
\ket{\hat{O}_\mathrm{A} \hat{O}_\mathrm{B}} = \hat{O}_\mathrm{A}\otimes \hat{I} \ket{\hat{O}_\mathrm{B}} = \hat{I} \otimes \hat{O}_\mathrm{B}^T \ket{\hat{O}_\mathrm{A}}.
\end{align}

With the above notations, we represent the regularized OTOC. We now assume that $\hat{O}_A$ and $\hat{O}_B$ are Hermitian. We define the Heisenberg representation of $\hat{O}_B$ for the complex-time evolution as 
$
\hat{O}_B(\beta,t)\coloneqq \mathrm{e}^{-\qty(\frac{\beta}{4}-\mathrm{i}t)\hat{H}}\hat{O}_B\mathrm{e}^{-\qty(\frac{\beta}{4}+\mathrm{i}t)\hat{H}}/\sqrt{Z}
$.
The OTOC is then written as
\begin{align}
\begin{aligned}
F(t)&=\tr[\hat{O}_B(\beta,t)\hat{O}_A  \hat{O}_B(\beta,t)\hat{O}_A] \\
&=\braket{\hat{O}_A\hat{O}_B(\beta,t)}{\hat{O}_B(\beta,t)\hat{O}_A} \\
&=\bra{\hat{O}_A}\hat{O}_B(\beta,t)\otimes \hat{O}_B(\beta,t) ^*\ket{\hat{O}_A}.
\end{aligned}
\end{align}
Using the cyclic property of the trace, we also obtain
\begin{align}
\begin{aligned}
F(t) &=\tr[\hat{O}_A \hat{O}_B(\beta,t)\hat{O}_A\hat{O}_B(\beta,t)] \\
&=\braket{\hat{O}_B(\beta,t) \hat{O}_A}{\hat{O}_A\hat{O}_B(\beta,t)}\\
& = \bra{\hat{O}_B(\beta,t)}\hat{O}_A\otimes \hat{O}_A^*\ket{\hat{O}_B(\beta,t)}.
\end{aligned}
\end{align}

We now prove inequality (\ref{eq:OTOCinequality}). A subsystem of A is denoted as $\mathrm{A}^\prime$. A basis operator $\hat{O}_\mathrm{A}$ can be decomposed into a basis operator of the subsystem $\mathrm{A}^\prime$ and that of $\mathrm{A}\setminus \mathrm{A}'$ as
\begin{align*}
& \hat{O}_\mathrm{A}=\hat{O}_{\mathrm{A}\setminus \mathrm{A}'} \hat{O}_{\mathrm{A}'}, \\
& [\hat{O}_{\mathrm{A}\setminus \mathrm{A}'},\hat{O}_{\mathrm{A}'}]=0.
\end{align*}
With this decomposition, we obtain~\cite{Xu}
\begin{align}
\nonumber
&\tr[\hat{O}_\mathrm{B}(\beta,t){\hat{O}_{\mathrm{A}}} \hat{O}_\mathrm{B}(\beta,t)\hat{O}_{\mathrm{A}}]\\&=\tr[\hat{O}_\mathrm{B}(\beta,t){\hat{O}_{\mathrm{A}'}} \hat{O}_\mathrm{B}(\beta,t)\hat{O}_{\mathrm{A}'}] \nonumber
\\&\;\;\;\;\;\;\;\; +\frac{1}{2}\tr[\hat{O}_{\mathrm{A}'}[\hat{O}_{\mathrm{A}\setminus \mathrm{A}'},\hat{O}_\mathrm{B}(\beta,t)]\hat{O}_{\mathrm{A}'}[\hat{O}_{\mathrm{A}\setminus \mathrm{A}'},\hat{O}_\mathrm{B}(\beta,t)]].
\end{align}
We then calculate the OTOC for the subsystems A and B as
\begin{align}
\label{eq:OTOC_AB_SUB_DECOMP}
\begin{aligned}
&F_{\mathrm{A}\mathrm{B}}(t)
\\
=&
\tr[\hat{O}_\mathrm{B}(\beta,t){\hat{O}_{\mathrm{A}'}} \hat{O}_\mathrm{B}(\beta,t)\hat{O}_{\mathrm{A}'}]
\\
&+
\frac{1}{2}\tr[\hat{O}_{\mathrm{A}'}[\hat{O}_{\mathrm{A}\setminus \mathrm{A}'},\hat{O}_\mathrm{B}(\beta,t)]\hat{O}_{\mathrm{A}'}[\hat{O}_{\mathrm{A}\setminus \mathrm{A}'},\hat{O}_\mathrm{B}(\beta,t)]]  \\
=&F_{\mathrm{A}'\mathrm{B}}(t)-\frac{1}{2}\tr[\hat{O}_\mathrm{A}'[\hat{O}_{\mathrm{A}\setminus \mathrm{A}'},\hat{O}_\mathrm{B}(\beta,t)]^\dagger \hat{O}_\mathrm{A}'[O_{\mathrm{A}\setminus \mathrm{A}'},\hat{O}_\mathrm{B}(\beta,t)]]\\
=&F_{\mathrm{A}'\mathrm{B}}(t)-\frac{1}{2}\braket{[\hat{O}_{\mathrm{A}\setminus \mathrm{A}'},\hat{O}_\mathrm{B}(\beta,t)]\hat{O}_{\mathrm{A}'}}{\hat{O}_{\mathrm{A}'}[\hat{O}_{\mathrm{A}\setminus \mathrm{A}'},\hat{O}_\mathrm{B}(\beta,t)]}\\
=& F_{\mathrm{A}'\mathrm{B}}(t) -\frac{1}{2}\bra{[\hat{O}_{\mathrm{A}\setminus \mathrm{A}'},\hat{O}_\mathrm{B}(\beta,t)]} \hat{O}_{\mathrm{A}'}\otimes \hat{O}_{\mathrm{A}'}^*\ket{[\hat{O}_{\mathrm{A}\setminus \mathrm{A}'},\hat{O}_\mathrm{B}(\beta,t)]}.
\end{aligned}
\end{align}

Finally, we take the average over the basis operator $\hat{O}_{\mathrm{A}'}$ in the subsystem $\mathrm{A}^\prime$ while $\hat{O}_{\mathrm{A}\setminus \mathrm{A}'}$ being fixed, and obtain 
\begin{align}
\sum_{\hat{O}_{\mathrm{A}'} \in \mathcal{P_{\mathrm{A}'}}}\hat{O}_{\mathrm{A}'}\otimes \hat{O}_{\mathrm{A}'}^* =\frac{1}{2^{a'}} \ket{\hat{I}_{\mathrm{A}'}}\bra{\hat{I}_{\mathrm{A}'}},
\end{align}
where $\hat{I}_{\mathrm{A}'}$ is the identity operator for the subsystem $\mathrm{A}^\prime$. By taking this average, the second term in the most right-hand side of Eq.~(\ref{eq:OTOC_AB_SUB_DECOMP}) can be regarded as the expectation value of a reduced density operator $\rho_{\mathrm{A}'}=\frac{1}{2^{a'}} \ket{\hat{I}_{\mathrm{A}'}}\bra{\hat{I}_{A'}}$ with a state vector $\ket{[\hat{O}_{\mathrm{A}\setminus \mathrm{A}'},\hat{O}_\mathrm{B}(\beta,t)]}$. Since the reduced density operator is positive semidefinite, this term should be non-negative. Thus, we obtain 
\begin{align}
\sum_{\hat{O}_{\mathrm{A}'} \in \mathcal{P_{\mathrm{A}'}}}F_{\mathrm{A}\mathrm{B}}(t) \leq \sum_{\hat{O}_{\mathrm{A}'} \in \mathcal{P_{\mathrm{A}'}}} F_{\mathrm{A}'\mathrm{B}}(t).
\end{align}
By taking the average over $\hat{O}_\mathrm{B}(\beta,t)$ and $\hat{O}_{\mathrm{A}\setminus \mathrm{A}'}$, we obtain the following inequality for the averaged OTOC:
\begin{align}
\overline{F}_{\mathrm{A}\mathrm{B}}(t)\leq\overline{F}_{\mathrm{A}'\mathrm{B}}(t).
\end{align}
By the decomposition of the subsystem B, we can perform the same procedure and finally obtain Eq.~(\ref{eq:OTOCinequality}). 

\section{Canonical form of iMPO}
\label{ap:caninical_iMPO}
In parallel to the canonical form of iMPS defined in Ref.~\cite{Orus2008}, we define the canonical form of the iMPO in the following. For simplicity, we consider a one-dimensional lattice system composed of $d$-level local systems. We can define the iMPO with the single-site translational invariance symmetry by two tensors $\qty{\Gamma,\lambda}$, where $\Gamma$ is a complex-valued tensor with four indices and $\lambda$ is a real-valued tensor with a single index. The elements of these tensors are represented by $\Gamma^{ij}_{\alpha \beta}$, and $\lambda$ is represented as a diagonal matrix with non-negative elements $\lambda_\alpha$. While the indices $\alpha$ and $\beta$ represent virtual degrees of freedom (virtual bond indices) and run within $1,\cdots,\chi$, the indices $i$ and $j$ represent physical degrees of freedom (physical bond indices) and run within $1,\cdots,d$. 

In order to define the canonial form, we introduce the following tensors $R$ and $L$:
\begin{align}
R_{\qty(\alpha \alpha'),\qty(\beta \beta ')} & \coloneqq \sum_{i=1,j=1}\qty(\Gamma^{ij}_{\alpha \beta}\lambda_\beta)\qty(\Gamma^{ij}_{\alpha'\beta'}\lambda_{\beta'})^*, \\
L_{\qty(\alpha \alpha'),\qty(\beta \beta ')} & \coloneqq \sum_{i=1,j=1}\qty(\lambda_\alpha\Gamma^{ij}_{\alpha \beta})\qty(\lambda_{\alpha'}\Gamma^{ij}_{\alpha'\beta'})^*.
\end{align}
The condition for the iMPO $\qty{\Gamma,\lambda}$ to be the canonical form is written as
\begin{align}
\sum_{\beta,\beta'}R_{\qty(\alpha \alpha'),\qty(\beta \beta ')} \delta_{\beta \beta'} &= \eta\delta_{\alpha \alpha'}, \\
\sum_{\alpha,\alpha'}\delta_{\alpha \alpha'} L_{\qty(\alpha \alpha'),\qty(\beta \beta ')} & =\eta\delta_{\beta \beta'}.
\end{align}
With the canonical form, the R\'enyi-$n$ OSEE for the bipartition $\mathrm{A}=\qty{-\infty,\cdots,-1,0}$ and $\mathrm{B}=\qty{1,2,\cdots\infty}$ can be calculated as
\begin{align}
{S_A^{\mathrm{OSEE\:}(n)}} = \frac{1}{1-n}\ln\sum_{\alpha=1}^{\chi}\qty(\lambda_\alpha^2)^n.
\end{align}

\section{Numerical error}
\label{ap:numerical}
In this Appendix, we discuss numerical errors in our calculation. We show that the errors from the Trotter decomposition and the bond dimensions are sufficiently small.
\subsection{Imaginary-time evolution}
\begin{figure}[h]
	\centering
	\subfigure{\includegraphics[width=1\linewidth]{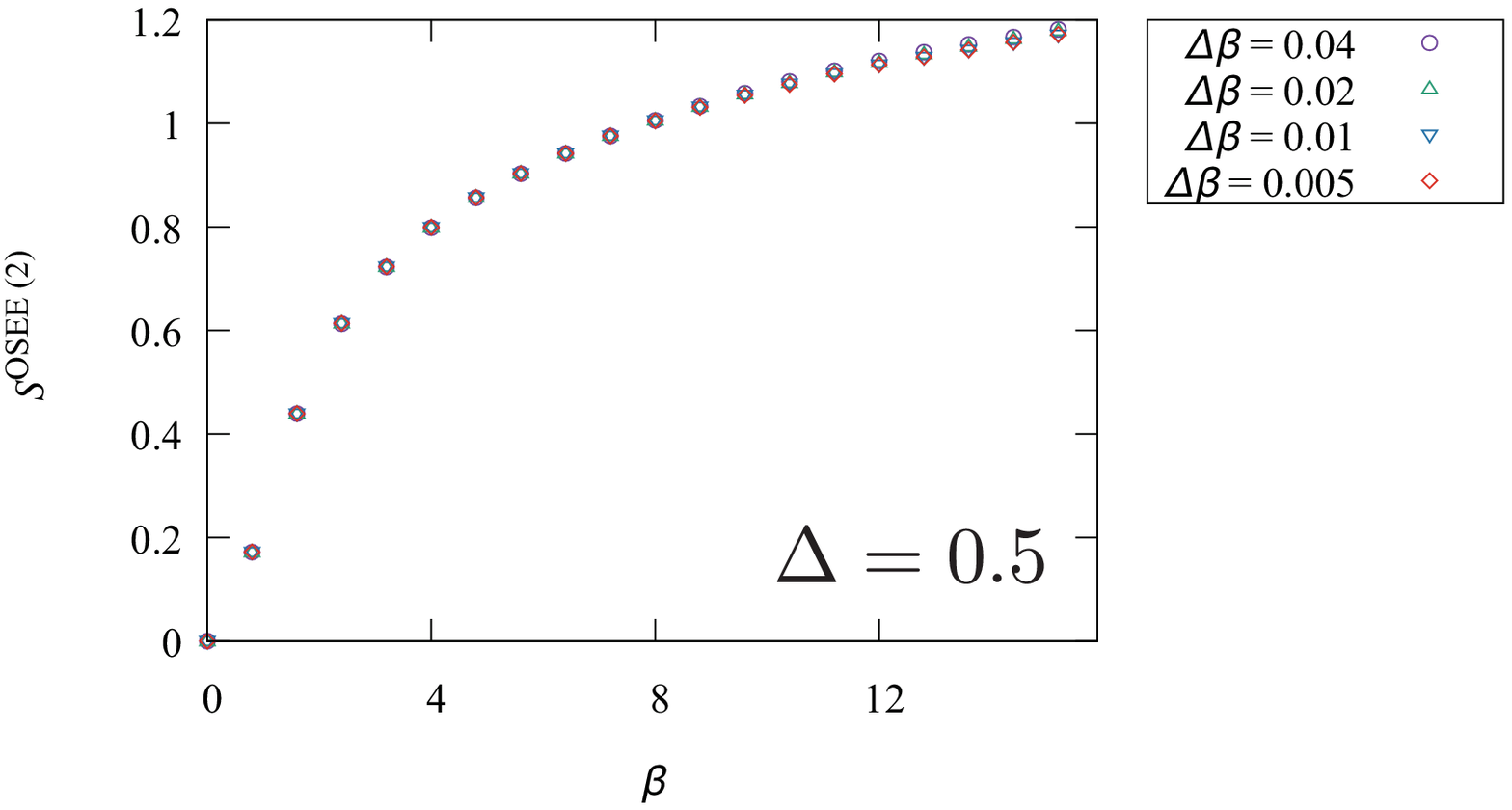}
		\label{fig:temperatureOSEE_bond512Jz05trotter_error.eps}}
	\hspace{20mm}
	\subfigure{\includegraphics[width=1\linewidth]{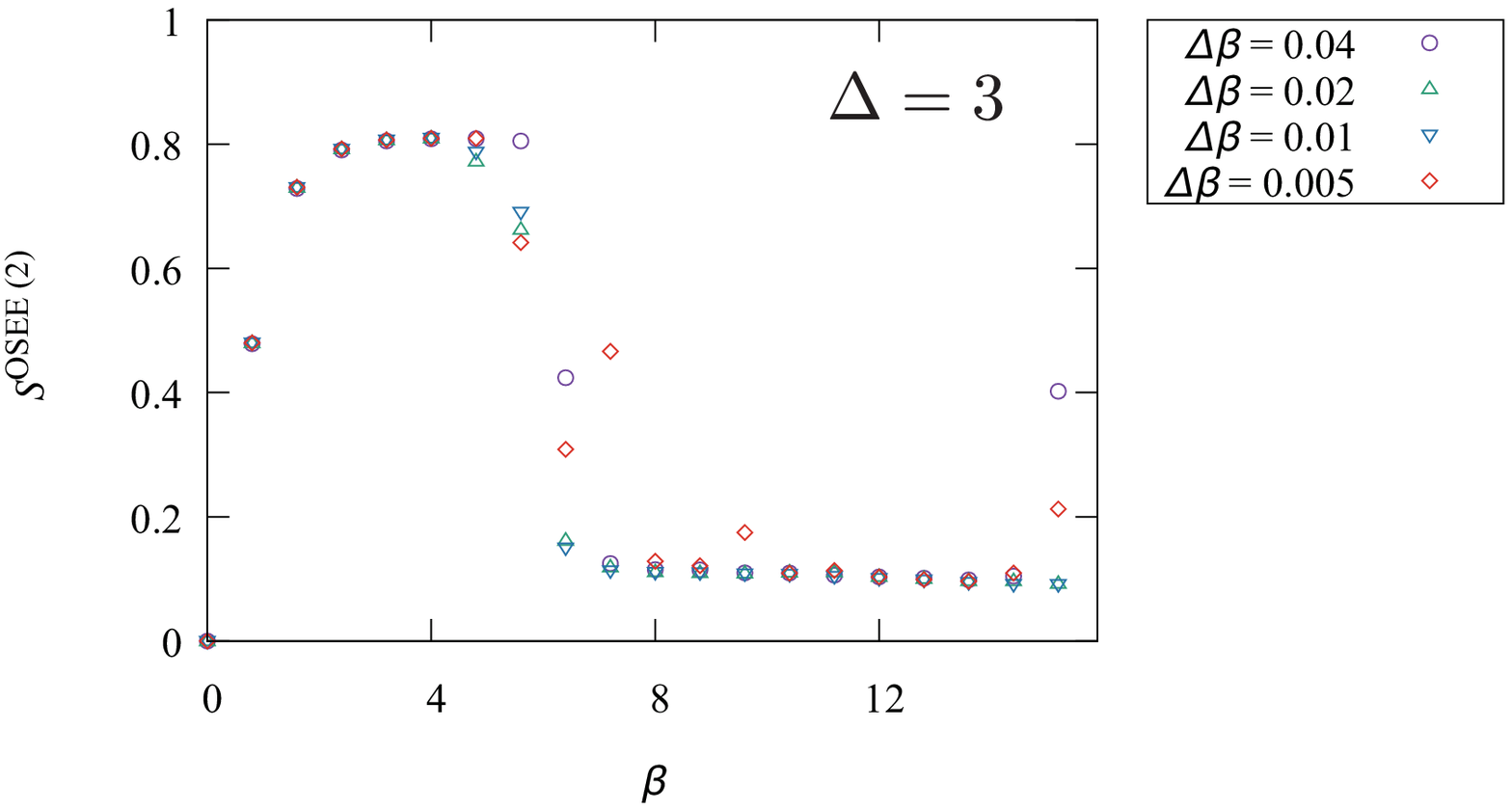}
		\label{fig:temperatureOSEE_bond512Jz3trotter_error.eps}}
	\caption{Imaginary-time evolution of the OSEE with $\Delta \beta = 0.04, 0.02,0.01, 0.005$. The bond dimension is taken as $\chi=512$. The upper figure is for the gapless region $\Delta=0.5$ and the lower one is for the gapped region $\Delta=3$.
	}
	\label{fig:temperatureOSEE_trotter_error}
\end{figure}

Figure~\ref{fig:temperatureOSEE_trotter_error} shows the dependence on the width in the Trotter decomposition of the imaginary-time evolution of the OSEE in the gapless region $\Delta=0.5$ and the gapped region $\Delta=3$. The width in the Trotter decomposition is taken as $\Delta \beta = 0.04, 0.02,0.01, 0.005$ and the bond dimension is fixed as $\chi =512$.

In the gapless region, the numerical errors from the Trotter decomposition are smaller than the point size. We estimate the numerical errors $\Delta S^{\mathrm{OSEE}}$ quantitatively from the difference between the maximum and minimum values for $\Delta \beta = 0.04, 0.02,0.01, 0.005$. We find that $\Delta S^{\mathrm{OSEE}} \sim 0.008$ at $\beta=8$ and $\Delta S^{\mathrm{OSEE}} \sim 0.01$ at $\beta=15.2$.

In the gapped region, while the numerical errors from the Trotter decomposition are smaller than the point size in the high-temperature region with small $\beta$, they are not as small in the low-temperature region with large $\beta$. We consider that this error in the low-temperature region comes from the artificial symmetry breaking in the ground state, which is discussed in Sec.~\ref{sec:imagtime}. In the high-temperature region, we estimate the error of the OSEE in the same manner as the gapless region and we obtain $\Delta S^{\mathrm{OSEE}} \sim 0.001$ at $\beta=4$.

\begin{figure}[th]
	\centering
	\subfigure{\includegraphics[width=1\linewidth]{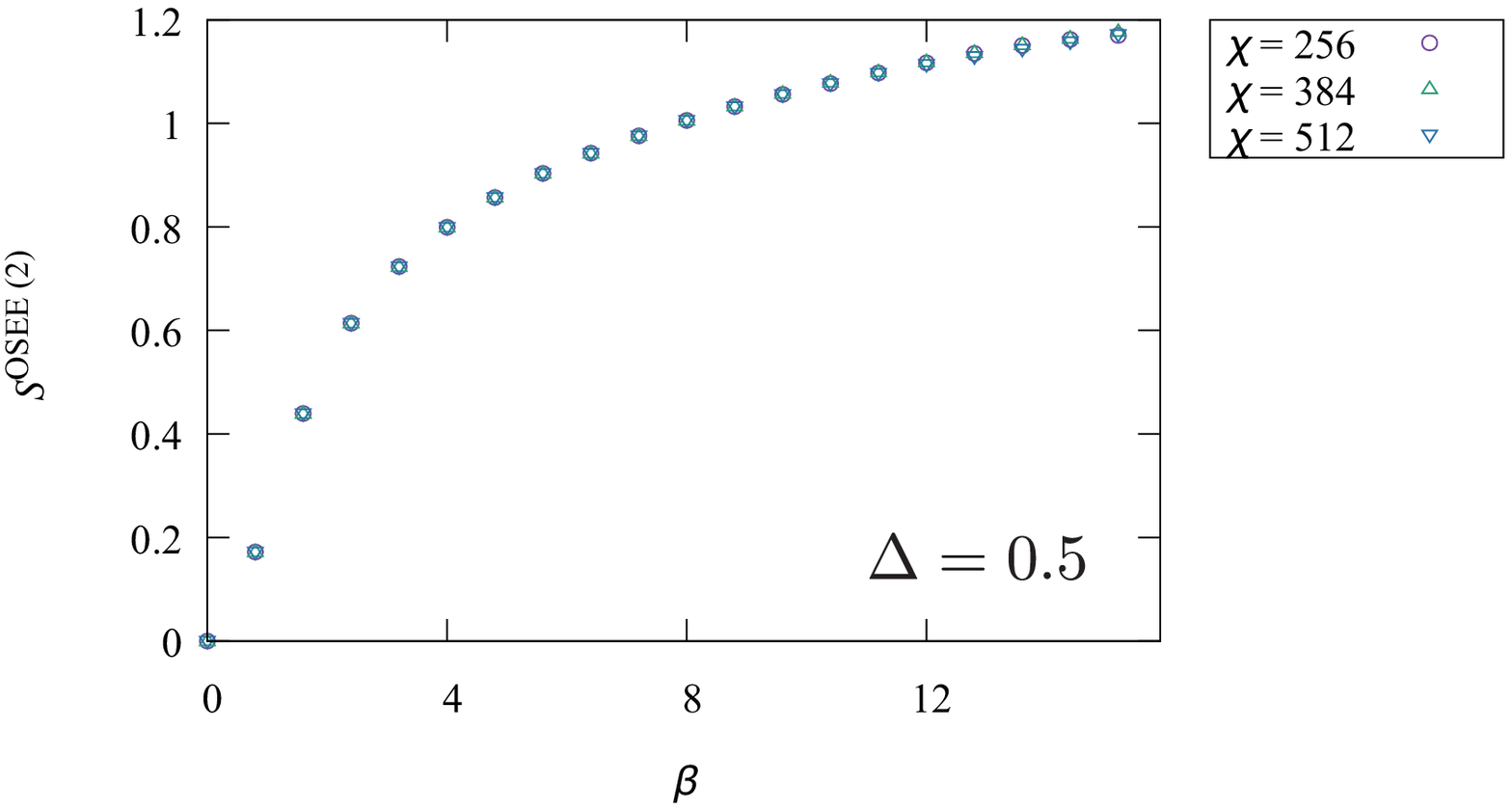}
		\label{fig:temperatureOSEE_trotter001Jz05bond_error.eps}}
	\hspace{20mm}
	\subfigure{\includegraphics[width=1\linewidth]{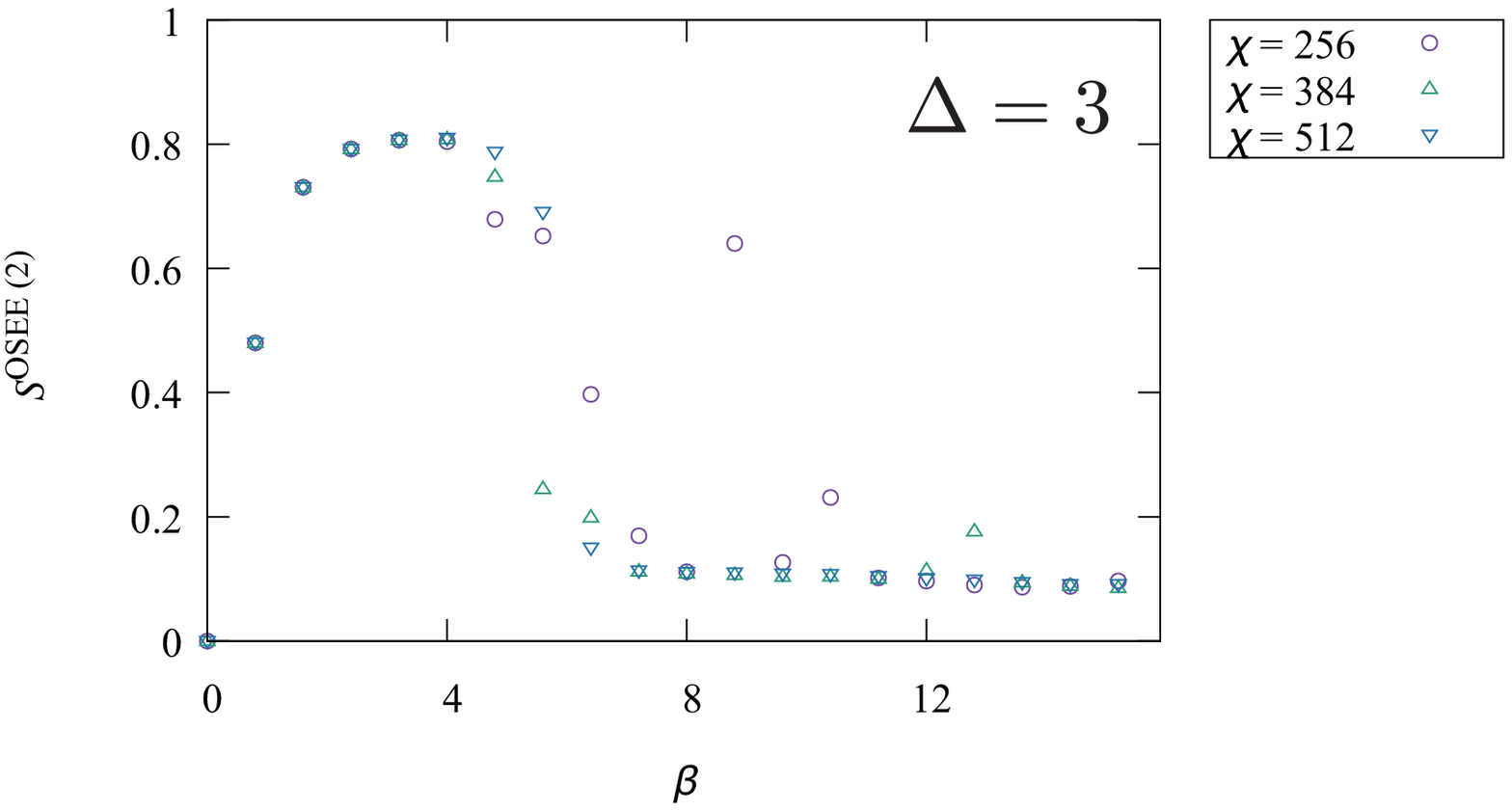}
		\label{fig:temperatureOSEE_trotter001Jz3bond_error.eps}}
	\caption{Imaginary-time evolution of the OSEE with $\chi=256,384,512$. The width in the Trotter decomposition is taken as $\Delta\beta = 0.01$. The upper figure is for the gapless region $\Delta=0.5$ and the lower one is for the gapped region $\Delta=3$.
}
	\label{fig:temperatureOSEE_bond_error}
\end{figure}

Figure~\ref{fig:temperatureOSEE_bond_error} shows the $\chi$-dependence of the imaginary-time evolution of the OSEE in the gapless region $\Delta=0.5$ and the gapped region $\Delta=3$. The bond dimension is taken as $\chi = 256,384,512$ and the width in the Trotter decomposition is fixed to $\Delta\beta = 0.01$.

In the gapless region, the numerical errors from the bond dimensions are smaller than the point size. We estimate the numerical errors $\Delta S^{\mathrm{OSEE}}$ in the same manner as the case of the Trotter decomposition. We find that $\Delta S^{\mathrm{OSEE}} \sim 0.009$ at $\beta=8$ and $\Delta S^{\mathrm{OSEE}} \sim 0.006$ at $\beta=15.2$.

In the gapped region, while the numerical errors from the bond dimensions are smaller than the point size in the high-temperature region, they are not small in the low-temperature region. This behavior is due to the artificial symmetry breaking in the ground state. We estimate the error of the OSEE and obtain $\Delta S^{\mathrm{OSEE}} \sim 0.006$ at $\beta=4$.
\subsection{Real-time evolution}

\begin{figure}[th]
	\centering
	\subfigure{\includegraphics[width=0.8\linewidth]{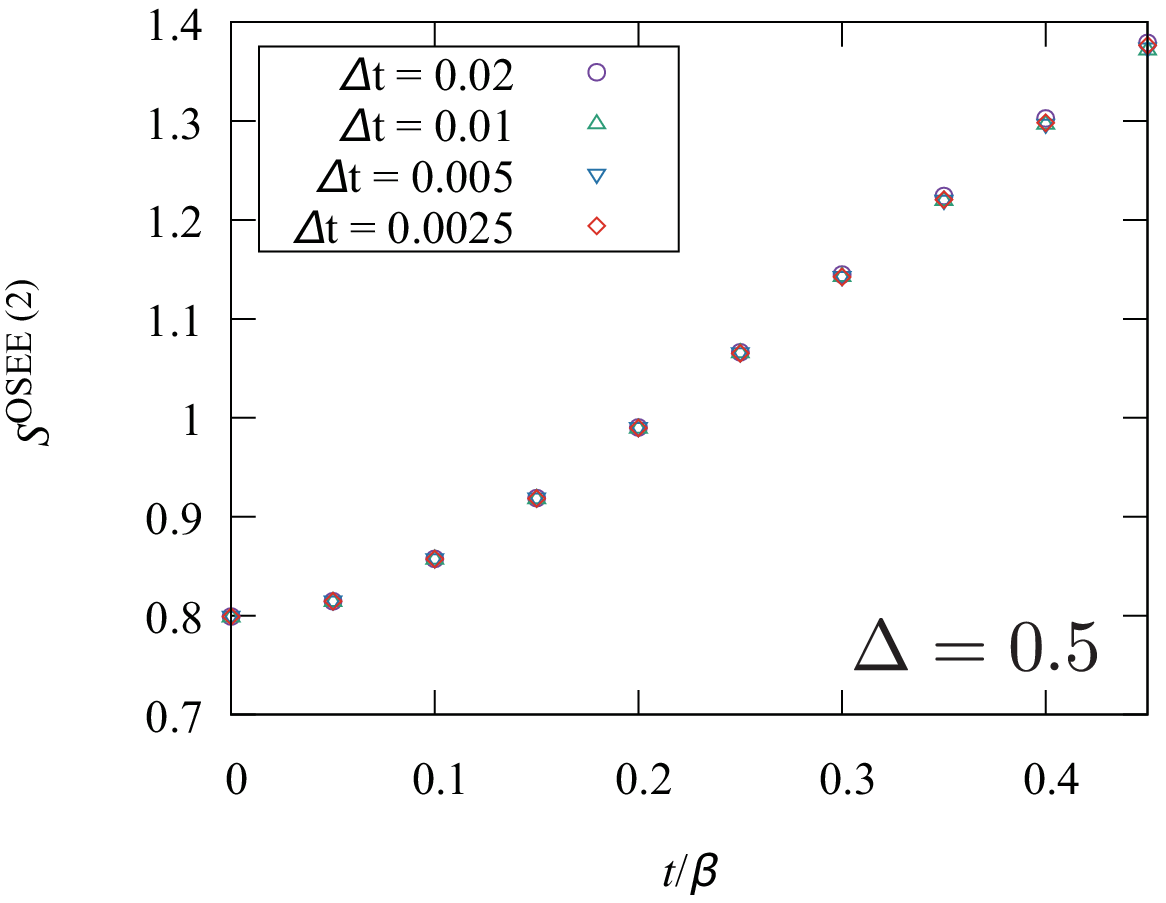}
		\label{fig:timeOSEE_bond512Jz05trotter_error.eps}}
	\hspace{20mm}
	\subfigure{\includegraphics[width=0.8\linewidth]{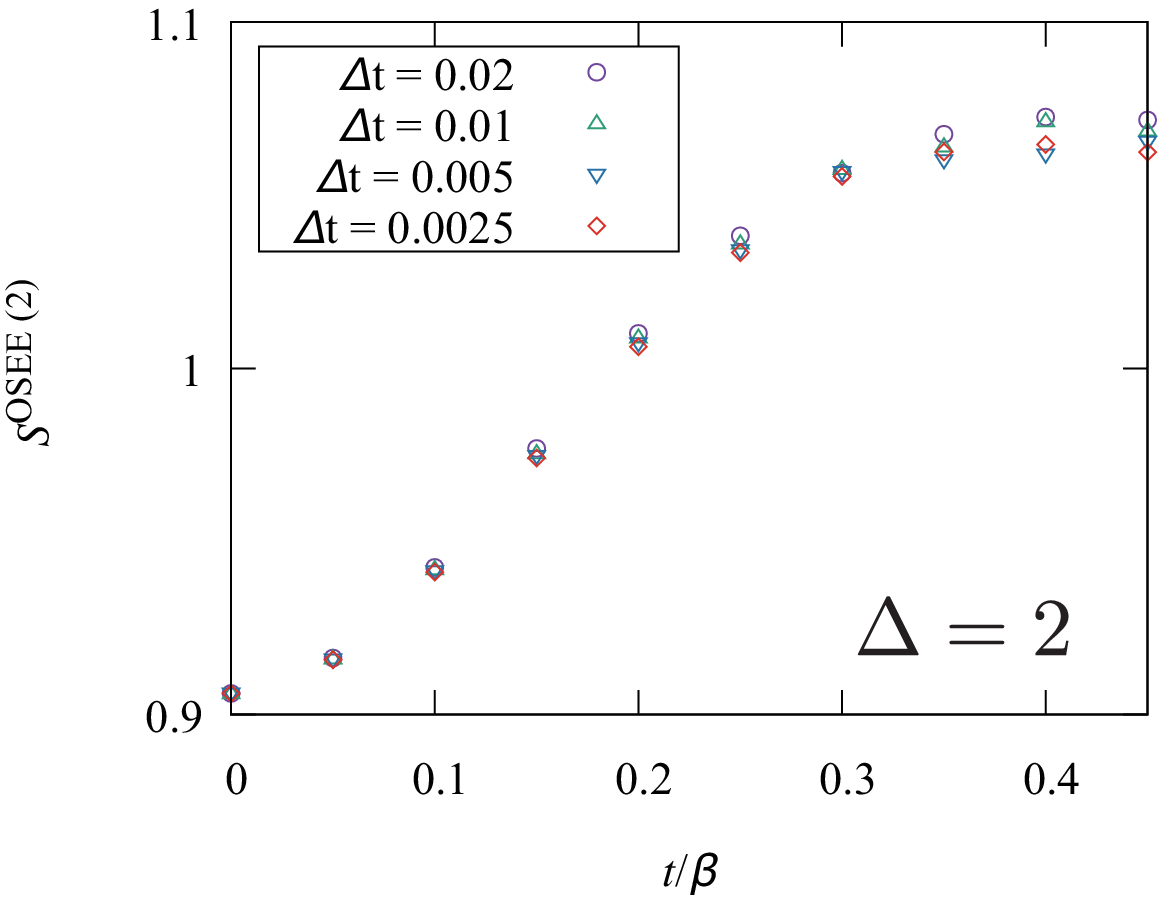}
		\label{fig:timeOSEE_bond512Jz3trotter_error.eps}}
	\caption{Real-time evolution of OSEE with $\Delta t =0.02,0.01,0.005,0.0025$. The inverse temperature is $\beta=4$ and the bond dimension is $\chi=512$. The upper figure is for the gapless region $\Delta=0.5$ and the lower one is for the gapped region $\Delta=2$.}
	\label{fig:timeOSEE_trotter_error}
\end{figure}

When we calculate the real-time evolution at finite temperature, we first calculate $\hat{U}(\beta,0)=\mathrm{e}^{-\frac{\beta}{4}\hat{H} }$ by the imaginary-time evolution and then calculate the real-time evolution to obtain $\hat{U}(\beta,t)=\mathrm{e}^{-\qty(\frac{\beta}{4}+\mathrm{i}t)\hat{H} }$. The width for the imaginary-time evolution is fixed as $\Delta \beta = 0.005$ and the inverse temperature is $\beta=4$. Figure~\ref{fig:timeOSEE_trotter_error} shows the real-time evolution of the OSEE with the width in the Trotter decomposition $\Delta t= 0.02,0.01,0.005,0.0025$.  We fix the bond dimension as $\chi=512$.

In the gapless region, the numerical errors from the Trotter decomposition are smaller than the point size. The numerical errors of the OSEE $\Delta S^{\mathrm{OSEE}}$ are estimated as $\Delta S^{\mathrm{OSEE}} \sim 0.006$ at $t/\beta=0.2$ and  $\Delta S^{\mathrm{OSEE}} \sim 0.006$ at $t/\beta=0.4$.

In the gapped region, though the numerical errors from the Trotter decomposition are not smaller than the point size, we have the result that the averaged OTOC decays faster in the gapless region than in the gapped region. The numerical errors are estimated as $\Delta S^{\mathrm{OSEE}} \sim 0.004$ at $t/\beta=0.2$ and $\Delta S^{\mathrm{OSEE}} \sim 0.01$ at $t/\beta=0.4$. When we consider $S^{\mathrm{OSEE}\:(n)}-S_0$ where $S_0$ is the OSEE at $t=0$, the difference between the OSEEs in the gapless and the gapped regions is approximately $0.08$ at $t/\beta=0.2$ and $0.34$ at $t/\beta=0.4$. The numerical errors of the OSEE from the Trotter decomposition are sufficiently smaller than the changes of $S^{\mathrm{OSEE}\:(n)}-S_0$ for our discussion.

\begin{figure}[th]
	\centering
	\subfigure{\includegraphics[width=0.8\linewidth]{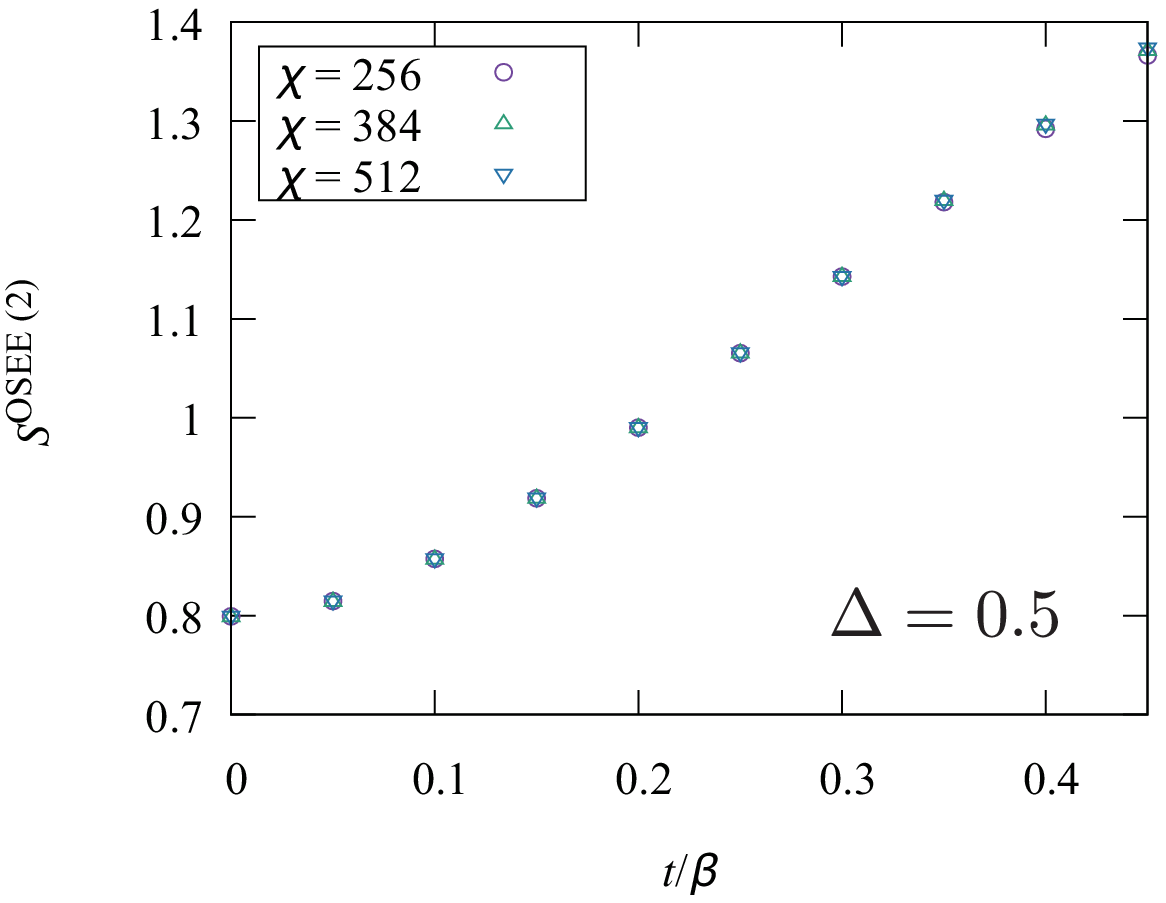}
		\label{fig:timeOSEE_trotter0005Jz05bond_error.eps}}
	\hspace{20mm}
	\subfigure{\includegraphics[width=0.8\linewidth]{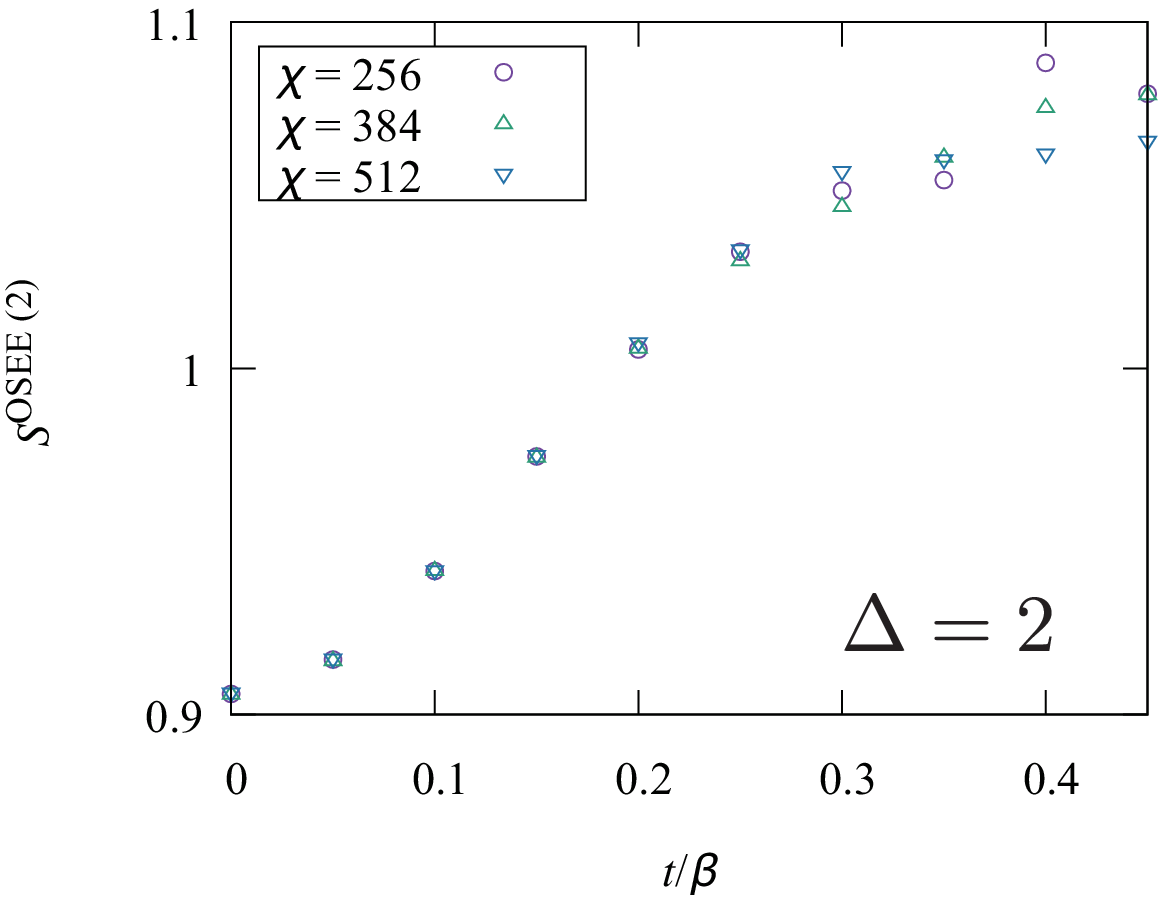}
		\label{fig:timeOSEE_trotter0005Jz2bond_error.eps}}
	\caption{Real-time evolution of the OSEE with $\chi= 256,384, 512$. The inverse temperature is $\beta=4$ and the width in the Trotter decomposition is $\Delta t =0.005$. The upper figure is for the gapless region $\Delta=0.5$ and the lower one is for the gapped region $\Delta=2$.}
	\label{fig:timeOSEE_bond_error}
\end{figure}

Figure~\ref{fig:timeOSEE_bond_error} shows the real-time evolution of the OSEE with $\chi =256,384, 512$.  We fix the width in the Trotter decomposition as $\Delta t = 0.005$.

In the gapless region, the numerical errors from the bond dimensions are smaller than the point size. We estimate the numerical errors $\Delta S^{\mathrm{OSEE}}$ and find that $\Delta S^{\mathrm{OSEE}} \sim 0.006$ at $t/\beta=0.2$ and $\Delta S^{\mathrm{OSEE}} \sim 0.006$ at $t/\beta=0.4$.

In the gapped region, though the numerical errors from the bond dimensions are not smaller than the point size, the result does not change because the errors are sufficiently smaller than the change of the OSEE. The numerical errors are estimated as $\Delta S^{\mathrm{OSEE}} \sim 0.008$ at $t/\beta=0.2$ and $\Delta S^{\mathrm{OSEE}} \sim 0.03$ at $t/\beta=0.4$. The difference between $S^{\mathrm{OSEE}\:(n)}-S_0$ in the gapless and the gapped regions is approximately $0.08$ at $t/\beta=0.2$ and $0.34$ at $t/\beta=0.4$. Thus, the numerical errors of the OSEE from the bond dimensions do not affect our conclusion.

\subsection{Role of scale invariance}
\begin{figure}[t]
	\centering
	\includegraphics[width=1\linewidth]{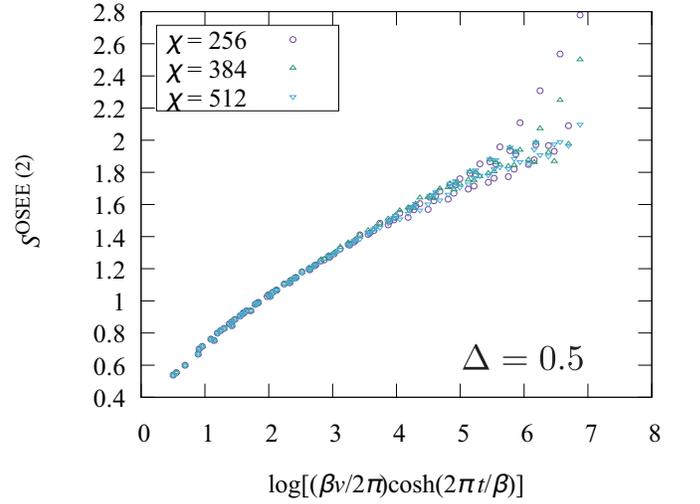}
	\caption{The $\ln L_{\mathrm{eff}}$ dependence of the OSEE of the complex-time evolution operator $\mathrm{e}^{-\left(\frac{\beta}{4}+\mathrm{i}t\right)\hat{H}}$ in the gapless region $\Delta = 0.5$ with $\chi=256,384,512$.}
	\label{fig:universalscalingOSEE_bond_all.eps}
\end{figure}

We calculate the real-time evolution of the R\'enyi-2 OSEE with $\chi=256,384,512$, $\Delta \beta = 0.005$ and $\Delta t = 0.005$. Figure~\ref{fig:universalscalingOSEE_bond_all.eps} shows the dependence of the OSEE on the effective length $L_\mathrm{eff}=\frac{\beta v_s}{2\pi}\cosh\frac{2 \pi}{\beta}t$. Each data series corresponds to the same $\chi$ and contains data with different temperatures $\beta=2,3,4,5,6$. When the effective length is sufficiently large as $\ln L_{\mathrm{eff}}\sim4$, the numerical errors from the bond dimensions become larger than the point size. 

\begin{figure}[t]
	\centering
	\includegraphics[width=1\linewidth]{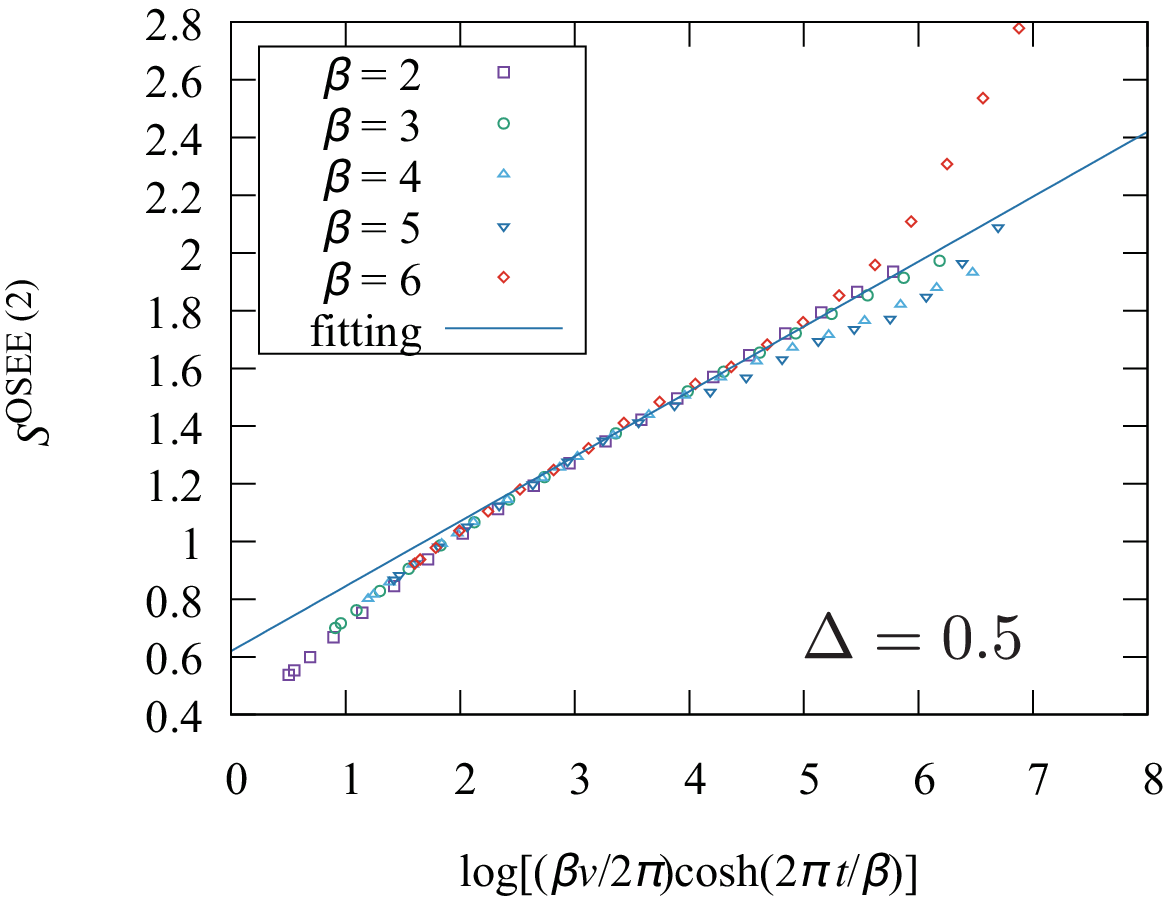}
	\caption{The $\ln L_{\mathrm{eff}}$ dependence of the OSEE of the complex-time evolution operator $\mathrm{e}^{-\left(\frac{\beta}{4}+\mathrm{i}t\right)\hat{H}}$ in the gapless region $\Delta = 0.5$ with $\chi=256$. The solid line represents the fitting result corresponding to the fitting range $[3,4]$ in Table~\ref{tab:c_a_256}.}
	\label{fig:timeOSEE_Jz0.5_256}
\end{figure}

\begin{figure}[t]
	\centering
	\includegraphics[width=1\linewidth]{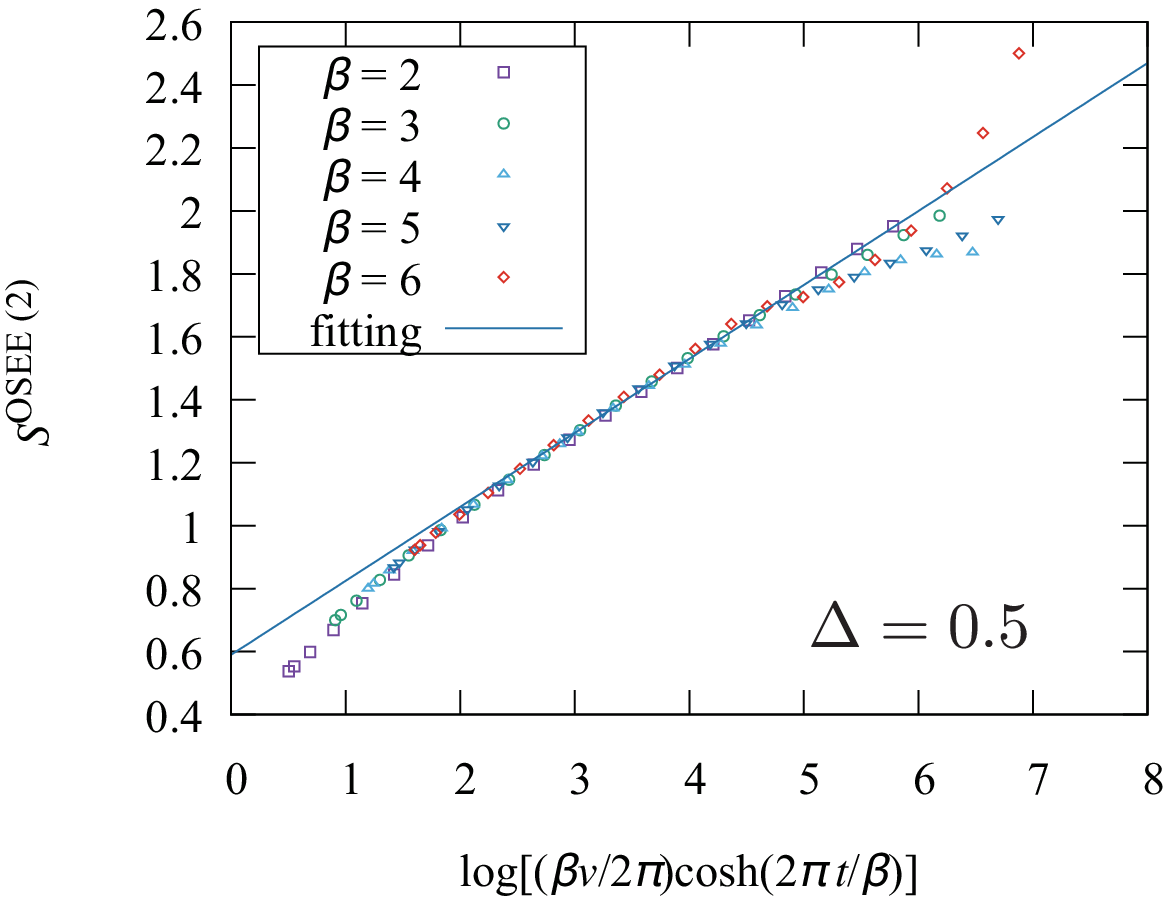}
	\caption{The $\ln L_{\mathrm{eff}}$ dependence of the OSEE of the complex-time evolution operator $\mathrm{e}^{-\left(\frac{\beta}{4}+\mathrm{i}t\right)\hat{H}}$ in the gapless region $\Delta = 0.5$ with $\chi=384$. The solid line represents the fitting result corresponding to the fitting range $[3,4]$ in Table~\ref{tab:c_a_384}.}
	\label{fig:timeOSEE_Jz0.5_384}
\end{figure}

Figures~\ref{fig:timeOSEE_Jz0.5_256} and \ref{fig:timeOSEE_Jz0.5_384} show $\ln L_{\mathrm{eff}}=\ln\qty(\frac{\beta v_s}{2\pi} \cosh(\frac{2\pi}{\beta}t))$ dependence of the R\'enyi-2 OSEE of the complex-time evolution operator $\mathrm{e}^{-\left(\frac{\beta}{4}+\mathrm{i}t\right)\hat{H}}$ with $\chi=256,384$. We obtain the data by calculating real-time evolution with fixing $\beta$ for each $\beta=2,3,4,5,6$. For each bond dimension, we fit the numerical data of all $\beta$'s with a fitting function ${S_A^{\mathrm{OSEE\:(2)}}}=\frac{c}{4}\ln L_{\mathrm{eff}}+a$, which comes from the 2D CFT. The fitting variables are $c$ and $a$. We adopt three fitting ranges $[2.5,4]$, $[3,4]$, and $[3.5,4]$, where the numerical errors from the bond dimensions are sufficiently small. Table~\ref{tab:c_a_256} and \ref{tab:c_a_384} show the results of the fitting for $\chi=256$ and $384$, respectively. The obtained value of $c$ is close to unity, which is consistent with the fact that the XXZ model with $\Delta = 0.5$ at low temperature is a quantum critical system effectively described by the 2D CFT.

\begin{table}[t]
	\label{tab:fitting_256}
	\centering
	\caption{Fitting results for the OSEE with a fitting function ${S_A^{\mathrm{OSEE\:(2)}}}=\frac{c}{4}\ln L_{\mathrm{eff}}+a$ with $\chi=256$.}
	\begin{tabular}{|l||c|c|}
		Fitting range & $c$ &  $a$ \\  \hline
		$[2.5,4]$& 0.939 $\pm$0.016 &0.584 $\pm$ 0.013   \\
		$[3,4]$& 0.90 $\pm$0.04 &0.62 $\pm$ 0.03   \\
		$[3.5,4]$& 0.89 $\pm$0.10 &0.62 $\pm$ 0.10   
		\label{tab:c_a_256}
	\end{tabular}
\end{table}
\begin{table}[t]
	\label{tab:fitting_384}
	\centering
	\caption{Fitting results for the OSEE with a fitting function ${S_A^{\mathrm{OSEE\:(2)}}}=\frac{c}{4}\ln L_{\mathrm{eff}}+a$ and $\chi=384$.}
	\begin{tabular}{|l||c|c|}
		FItting range & $c$ &  $a$ \\  \hline
		$[2.5,4]$& 0.969 $\pm$0.014 &0.5663 $\pm$ 0.011   \\
		$[3,4]$& 0.94 $\pm$0.03 &0.59 $\pm$ 0.02   \\
		$[3.5,4]$& 0.90 $\pm$0.08 &0.63 $\pm$ 0.07  
		\label{tab:c_a_384}
	\end{tabular}
\end{table}

\clearpage

\bibliography{MyCollection}
\end{document}